\documentclass[aps,prd,superscriptaddress,floatfix,nofootinbib]{revtex4}

\usepackage{verbatim}
\usepackage{mathtools}
\usepackage{ragged2e}
\usepackage{amsmath}
\usepackage{amsfonts}
\usepackage{amssymb}
\usepackage{graphicx}
\usepackage{slashed}
\usepackage{bm}
\usepackage{color}
\usepackage{epsf}
\usepackage{orcidlink}
\usepackage[parfill]{parskip}
\usepackage{hyperref}
\hypersetup{
  colorlinks   = true, %Colours links instead of ugly boxes
  urlcolor     = blue, %Colour for external hyperlinks
  linkcolor    = black, %Colour of internal links
  citecolor   = red %Colour of citations
}

\newcommand{\pI}{\mathtt{p}}
\newcommand{\qI}{\mathtt{q}}
\newcommand{\kI}{\mathtt{k}}

\newcommand{\modppI}{|\vec{\pI'}|} % modulus of the 3-vector p' in TRF-I

\newcommand{\eps}{\varepsilon}
\newcommand{\thetaG}{\theta_\gamma} % angle between 3-vectors q and q'
\newcommand{\modv}{|\vec{v}|} % modulus of the 3-velocity between TRF-I and TRF-II
\newcommand{\bq}{\bar{q}} % \bq = (q + q')/2
\newcommand{\bQ}{\bar{Q}} % \bQ^2 = -\bq^2
\newcommand{\bp}{\bar{p}} % \bp = (p + p')/2
\newcommand{\bxB}{\rho} % generalized Bjorken scale
\newcommand{\thetaL}{\theta_\ell} % polar angle of muon's 3-momentum in leptons' CM frame
\newcommand{\phiL}{\phi_\ell} % azimuthal angle of muon's 3-momentum in leptons' CM frame
\newcommand{\M}{\mathcal{M}} % amplitude
\newcommand{\K}{\kappa} 
\newcommand{\q}{\mathfrak{q}} % quark field
\newcommand{\cffh}{\mathcal{H}} % CFF H
\newcommand{\cffe}{\mathcal{E}} % CFF E
\newcommand{\cffht}{\mathcal{\widetilde{H}}} % CFF Htilde
\newcommand{\cffet}{\mathcal{\widetilde{E}}} % CFF Etilde
\newcommand{\Joplus}{\mathcal{J}^{(1)+}}
\newcommand{\Jt}{\mathcal{J}^{(2)}}
\newcommand{\Jofplus}{\mathcal{J}^{(1, 5)+}}
\newcommand{\Jtfplus}{\mathcal{J}^{(2, 5)+}}

\newcommand{\thetaS}{\theta_S}
\newcommand{\phiS}{\phi_S}
\newcommand{\vphiS}{\varphi_S}

\newcommand{\phiTrento}{\phi_{\mathrm{Trento}}}
\newcommand{\phiSTrento}{\varphi_{S, \mathrm{Trento}}}
\newcommand{\phiLBDP}{\phi_{\ell, \mathrm{BDP}}}
\newcommand{\thetaLBDP}{\theta_{\ell, \mathrm{BDP}}}
\newcommand{\OmegaLBDP}{\Omega_{\ell, \mathrm{BDP}}}
\newcommand{\ci}{c_\chi}
\newcommand{\si}{s_\chi}

\DeclareMathOperator\arctanh{arctanh}

\def\bea{\begin{eqnarray}}
\def\eea{\end{eqnarray}}
\def\beas{\begin{eqnarray*}}
\def\eeas{\end{eqnarray*}}
\def\beqas{\begin{eqnarray*}}
\def\eqas{\end{eqnarray*}}
\def\beq{\begin{equation}}
\def\eeq{\end{equation}}
\def\beqd{\begin{displaymath}}
\def\eeqd{\end{displaymath}}
\def\eqd{\end{displaymath}}

\def\slashchar#1{\setbox0=\hbox{$#1$}
   \dimen0=\wd0
   \setbox1=\hbox{/} \dimen1=\wd1
   \ifdim\dimen0>\dimen1
      \rlap{\hbox to \dimen0{\hfil/\hfil}}
      #1
   \else\begin{eqnarray}
      \rlap{\hbox to \dimen1{\hfil$#1$\hfil}}
      /
   \fi}

\begin{document}

\title{Phenomenology of double deeply virtual Compton scattering in the era of  new experiments}

\author{K.~Deja\,\orcidlink{0000-0002-9083-2382}}
\affiliation{National Centre for Nuclear Research (NCBJ), 02-093 Warsaw, Poland}

\author{V.~Mart\'inez-Fern\'andez\,\orcidlink{0000-0002-0581-7154}}
\affiliation{National Centre for Nuclear Research (NCBJ), 02-093 Warsaw, Poland}
\author{B.~Pire\,\orcidlink{0000-0003-4882-7800}}
\affiliation{Centre de Physique Th\'eorique, CNRS, École Polytechnique, I.P. Paris, 91128 Palaiseau, France  }

\author{P.~Sznajder\,\orcidlink{0000-0002-2684-803X}}
\affiliation{National Centre for Nuclear Research (NCBJ), 02-093 Warsaw, Poland}

\author{J.~Wagner\,\orcidlink{0000-0001-8335-7096}}
\affiliation{National Centre for Nuclear Research (NCBJ), 02-093 Warsaw, Poland}

\date{\today}

\begin{abstract}
We revisit the phenomenology of the deep exclusive electroproduction of a lepton pair, i.e. double deeply virtual Compton scattering (DDVCS), in view of new experiments planned in the near future. The importance of DDVCS in the reconstruction of generalized parton distributions (GPDs) in their full kinematic domain is emphasized. Using Kleiss-Stirling spinor techniques, we provide the leading order complex amplitudes for both DDVCS and Bethe-Heithler sub-processes. Such a formulation turns out to be convenient for practical implementation in the PARTONS framework and EpIC Monte Carlo generator that we use in simulation studies. 
\end{abstract}

\maketitle

\section{Introduction}

\begin{figure}[!ht]
    \centering
    \includegraphics{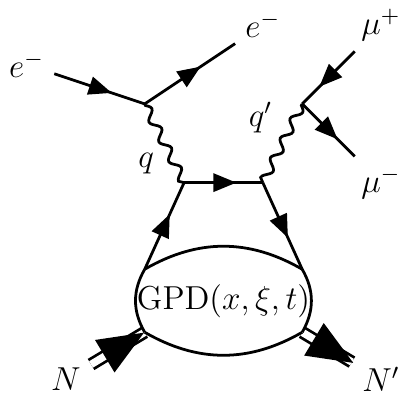}
    \includegraphics{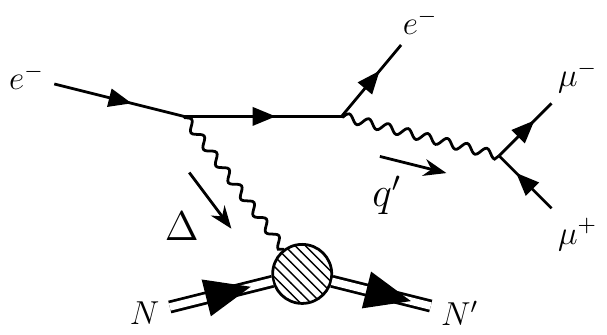}
    \includegraphics{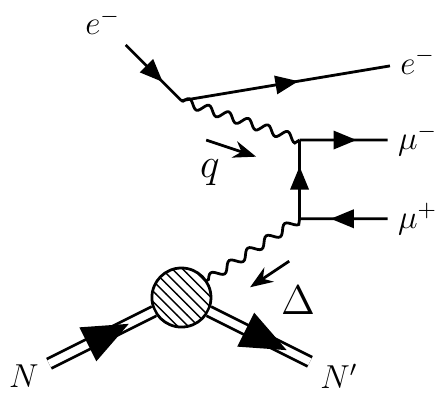}
    \caption{(from left to right) The double deeply virtual Compton scattering (DDVCS) process at leading order  and the two types of Bethe-Heitler processes, which contribute to the electroproduction of a lepton pair. Complementary crossed diagrams  are not shown in this figure.}
    \label{figure::RelevantDiagrams}
\end{figure}

The exclusive electroproduction of a lepton pair,
\begin{equation}
    e(k) + N(p)  \to e'(k') + N'(p') + \mu^+(\ell_+) + \mu^-(\ell_-) \,,
    \label{reaction}
\end{equation}
receives contributions from two sub-processes having the same initial and final states. One of them is purely QED Bethe-Heitler process (BH), while the other one is double deeply virtual Compton scattering (DDVCS):
\begin{equation}
    \gamma^*(q) + N(p)  \to \gamma^*(q') + N'(p') \,.
    \label{reaction2}
\end{equation}
 In the generalized Bjorken limit the amplitude of the latter is known \cite{Mueller:1998fv} to factorize into perturbatively calculable coefficient functions and generalized parton distributions (GPDs) that unravel the three-dimensional structure of the nucleon \cite{Diehl:2003ny,Belitsky:2005qn}. Conditions necessary to factorize DDVCS amplitude are the existence of a large scale, which may be either $Q^2 = -q^2 = -(k-k')^2$ or $Q'^2 = q'^2 = (\ell_+ +\ell_-)^2$, and relatively small squared four-momentum transfer to the nucleon, $-t = -(p'-p)^2$. Diagrams for both BH and DDVCS sub-processes are shown in Fig.~\ref{figure::RelevantDiagrams}. 

The DDVCS process has never been measured so far. Its importance comes from the fact that it is a most promising source of experimental information on GPDs. This can easily be seen in the fact that at the lowest order (LO) the deeply virtual Compton scattering (DVCS) \cite{Belitsky_2002} and timelike Compton scattering (TCS) \cite{Berger:2001xd} amplitudes, which are limits of DDVCS at $Q'^2 \to 0$ (DVCS) and $Q^2 \to 0$ (TCS), depend only on the $x=\xi$ domain, where $x$ is the average parton momentum, and $\xi$ describes the longitudinal momentum transfer. For the ongoing discussion at NLO see Ref.~\cite{Bertone:2021yyz}. The existence of two large scales in DDVCS, $Q^2$ and $Q'^2$, makes this process sensitive to the $x\neq\xi$ region, already at LO. We note here for completeness that another promising source of GPD information in the unexplored $x\neq\xi$ domain is lattice-QCD \cite{Bhattacharya:2022aob, Joo:2020spy}. 

The importance of DDVCS has  already been discussed some twenty years ago \cite{Belitsky:2002tf,guidal2003}. Although a quite detailed study of the phenomenological peculiarities of DDVCS already exists \cite{Belitsky:2003fj}, it is deemed appropriate to revisit the promises of this process, as it will be studied with great care in the near future at both fixed target facilities \cite{Chen:2014psa,Camsonne:2017yux,Zhao:2021zsm} and electron-ion colliders \cite{AbdulKhalek:2021gbh, Anderle:2021wcy}. Thanks to the use of Kleiss-Stirling techniques \cite{Kleiss:1984dp, Kleiss:1985yh}, we are allowed to present both the complete BH and the leading twist (LT) leading order DDVCS amplitudes as complex quantities. Such a formulation is convenient for practical implementation. We use the open-source PARTONS framework \cite{Berthou:2015oaw} to perform the numerical evaluation of DDVCS observables. Results of our work have also been implemented in the EpIC Monte Carlo generator \cite{Aschenauer:2022aeb} for a tentative estimate of DDVCS measurability in future experiments. 

The content of this article is as follows. In Sect.~\ref{sectKin} the kinematics of DDVCS is reviewed. In Sect.~\ref{sectAmp} we calculate the scattering amplitudes  for both Bethe-Heitler and leading order (in $\alpha_s$) DDVCS sub-processes using Kleiss-Stirling spinor techniques \cite{Kleiss:1984dp,Kleiss:1985yh}. In Sect.~\ref{sectLimits} both $Q^2 \to 0$ and $Q'^2 \to 0$ limits of our calculation are compared with independent results for TCS and DVCS. In Sect.~\ref{sectObs} we discuss a few specific observables for DDVCS, while in Sect.~\ref{sectMC} we present results obtained in our Monte Carlo study, which give some hints on the measurability of this process. Section~\ref{sectConc} provides the summary of this article.

\section{Kinematics}
\label{sectKin}

\subsection{Reference frames and momenta parametrization}\label{sec::frames}

\begin{figure}[!ht]
    \centering
    \includegraphics[trim={0 0 0 9px},clip]{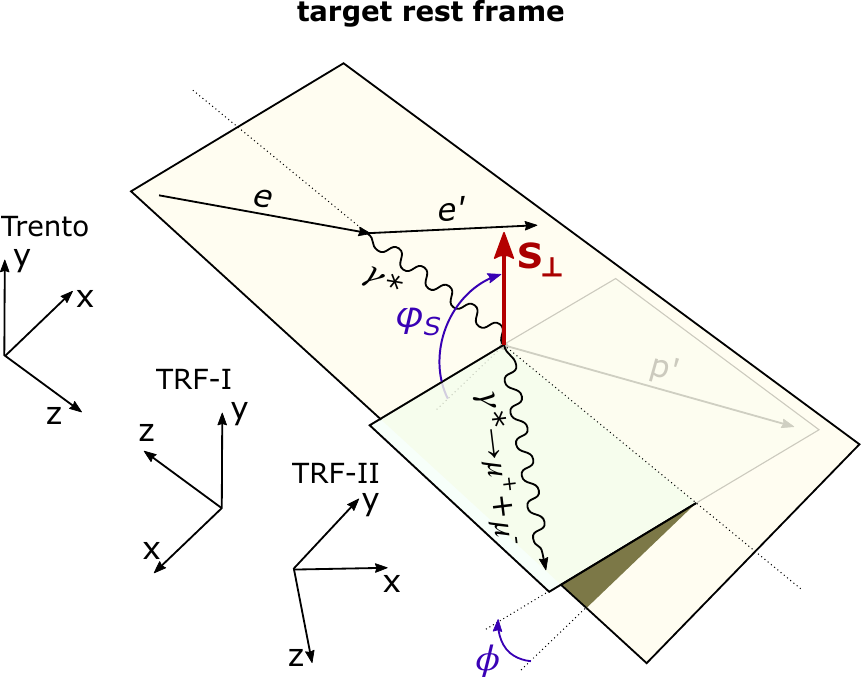}
    \includegraphics[trim={0 0 0 9px},clip]{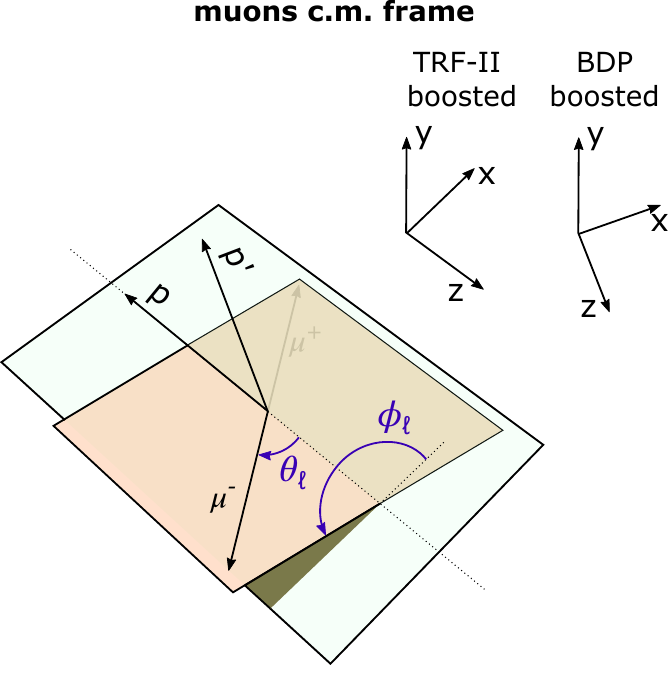}
    \caption{Kinematics of DDVCS process with the depiction of coordinate systems discussed in this article. (left)~Frame with initial proton at rest and transverse component of target polarisation vector, $\vec{S}_\perp$, defined with respect to the incoming virtual photon. (right)~Produced lepton pair center of mass frame.}
    \label{figure::DDVCSsetUp}
\end{figure}

In this section we describe the kinematics of the DDVCS process  \eqref{reaction}. The core of our evaluation is done in reference frames that coincide with those used in Ref. \cite{Belitsky:2002tf}. This choice allows us to stay consistent with the literature on the DDVCS topic, but also, in some cases, to facilitate comparison of obtained results. Relations to the Trento frame \cite{trento}, which is typically used to describe the DVCS process, and the frame popular in TCS analyses \cite{Berger:2001xd} are discussed in the following. The kinematical quantities are depicted in Fig. \ref{figure::DDVCSsetUp}.

We start the discussion by defining the ``target rest frame I'' (TRF-I), where the initial-state proton stays at rest, and where the $z-$axis is opposite to the incoming photon momentum. In this frame the four-momenta of the initial-state proton and of the incoming photon are:
\beq
\pI^\mu = (M, \vec{0}),\quad \qI^\mu = (\qI^0, 0, 0, \qI^3),
\eeq
where $M$ is the proton mass and $\qI^3 < 0$. With four-momenta of incoming electron, scattered electron, initial- and final-state protons denoted by $\kI$, $\kI'$, $\pI$, $\pI'$, respectively, we may write the incoming virtuality $Q^2$, Mandelstam variable $t$ and  Bj\"orken variable $x_B$ as:
\beq
Q^2 = -\qI^2 = -(\kI-\kI')^2,\quad t = \Delta^2 = (\pI' - \pI)^2,\quad x_B = Q^2/(2\pI\qI) \,.
\eeq
Four-momenta components of incoming virtual photon and final-state proton expressed in terms of these invariants read:
\begin{gather}\label{q0q3}
\qI^0 = \frac{Q}{\eps},\quad \qI^3 = -\frac{Q}{\eps}\sqrt{1 + \eps^2},\\
\pI^{\prime\mu} = \left( M - \frac{t}{2M},\, \modppI\sin\theta_N\cos\phi,\, \modppI\sin\theta_N\sin\phi,\, \modppI\cos\theta_N \right) \,.
\end{gather}
Here, $\eps = 2x_BM/Q$, while $\phi$ and $\theta_N$ are azimuthal and polar angles, respectively, given with respect to TRF-I axes. The angle $\theta_N$ is fixed by the momentum conservation and expressed in the following. We also define the angle $\vphiS$  between the $x$-axis and the transverse component of the initial proton polarization vector with respect to the incoming photon direction.

The momentum of the final-state proton is:
\beq
\modppI = \sqrt{-t\left( 1 - \frac{t}{4M^2} \right)}\,.
\eeq
The four-momentum of the outgoing photon with timelike virtuality $\qI'^2 = Q'^2$ is:
\beq 
\qI'^\mu = \qI'^0(1, \vec{v}),
\eeq
where
\beq\label{qp0v}
\qI'^0 = \frac{Q}{\eps} + \frac{t}{2M},\quad \modv = \sqrt{1 - \left( \frac{Q'}{\qI'^0} \right)^2}\,.
\eeq
The angle $\theta_N$ in terms of $\pI'$ and $\qI'$ reads:
\beq\label{cosN}
\cos\theta_N = -\frac{\eps^2(Q^2 + Q'^2 - t) - 2x_B t}{4x_B M \modppI \sqrt{1 + \eps^2}}\,.
\eeq
Four-momentum of the incoming electron in the massless limit is:
\begin{equation}
    \kI^\mu = E(1,\, \sin\theta_e,\, 0,\, \cos\theta_e)\,,
\end{equation}
where $E$ stands for the electron beam energy, while $\theta_e$ is the polar angle of electron's momentum in TRF-I. This angle is given by:
\begin{equation}
    \cos\theta_e = -\frac{1}{\sqrt{1+\eps^2}}\left( 1 + \frac{y\eps^2}{2} \right)\,,
\end{equation}
where $y = \pI\qI/(\pI\kI) = Q/(\eps E)$ is the inelasticity variable.

It is convenient to describe the produced leptons in their center-of-mass frame. In this work, again in correspondence to Ref. \cite{Belitsky:2002tf}, we define a new coordinate system called TRF-II and then, if required, we boost the relevant four-momenta along the direction of the outgoing photon momentum. TRF-II is rotated with respect to TRF-I by the angle $\thetaG$ between $z$-axis in TRF-I and $\vec{\qI'}$. This angle is given by:
\begin{equation}
    \cos\thetaG = -\frac{\eps (Q^2 - Q'^2 + t)/2  + Q\qI'^0}{Q\modv \qI'^0\sqrt{1+\eps^2}}\,.
\end{equation}
The transformation between TRF-I and TRF-II, without any boost, can be described by the Lorentz transformation matrix:
\begin{equation}
    \mathcal{R}_{\rm II\leftarrow I} = \left(\begin{array}{c|c}
        1 & 0_{1\times 3} \\
        \cline{1-2}
        0_{3\times 1} & (R_{\rm II\leftarrow I})_{3\times 3}
    \end{array}\right),\quad 
    (R_{\rm II\leftarrow I})_{3\times 3} = 
    \begin{pmatrix}
    -c_\gamma c_{\phi} & -c_\gamma s_{\phi} & -s_\gamma \\
    s_{\phi} & -c_{\phi} & 0 \\
    -s_\gamma c_{\phi} & -s_\gamma s_{\phi} & c_\gamma
    \end{pmatrix}\,,
\end{equation}
where short-hands $c_\gamma = \cos\thetaG$, $c_{\phi} = \cos\phi$, and likewise for sines, are used. With this transformation the momenta displayed previously in TRF-I in TRF-II read:
\begin{align}
    q'^\mu & = \qI'^0(1, 0, 0, \modv) \label{q'-TRFII} \,,\\
    q^\mu & = (\qI^0, -s_\gamma \qI^3, 0, c_\gamma \qI^3) \label{q-TRFII} \,,\\
    k^\mu & = E(1, -s_e c_\gamma c_\phi - c_e s_\gamma, s_e s_\phi, -s_e s_\gamma c_\phi + c_e c_\gamma ) \label{k-TRFII} \,,\\
    p^\mu & = \pI^\mu \,, \label{p-TRFII}
\end{align}
where $\qI^0, \qI^3$ and $\qI'^0$ are given in Eqs.~(\ref{q0q3}) and (\ref{qp0v}), and where $s_e = \sin\theta_e, c_e = \cos\theta_e$.

Since muon mass effects are of order $(m_\ell/Q')^2$ we neglect them by making four-momenta of the produced leptons light-like. In the center-of-mass frame of the lepton pair, i.e. in TRF-II frame boosted along its $z$-axis, these four-momenta are given by:
\begin{equation}
    \ell_{\mp}^\mu = \frac{Q'}{2}(1, \pm\vec{\beta}),\quad \vec{\beta} = (\sin\thetaL\cos\phiL, \sin\thetaL\sin\phiL, \cos\thetaL)\,.
\end{equation}
To revert the boost we use the velocity $\modv$ given in Eq.~(\ref{qp0v}), which gives:
\begin{equation}
    \ell_-^\mu = \left( \frac{1}{2}\qI'^0(1+\modv\cos\thetaL), \frac{1}{2}Q'\sin\thetaL\cos\phiL, \frac{1}{2}Q'\sin\thetaL\sin\phiL, \frac{1}{2}\qI'^0(\modv +  \cos\thetaL) \right)
\end{equation}
and $\ell_+$ is obtained by the substitution $(\phiL, \thetaL)\rightarrow (\pi+\phiL, \pi-\thetaL)$.

\subsection{Additional variables}

We introduce the following combinations of four-momenta:
\begin{equation}
    \Delta = p'-p = q-q',\quad \bq = \frac{q+q'}{2},\quad \bp = \frac{p+p'}{2}\,,
\end{equation}
which are used to define the skewness $\xi$ and ``generalized'' Bj\"orken variable $\bxB$ \cite{Diehl:2003ny}:
\begin{gather}
    \xi = -\frac{\Delta\bq}{2\bp\bq} = \frac{Q^2+Q'^2}{2Q^2/x_B - Q^2 - Q'^2 + t}\,,\\ \bxB = \frac{\bQ^2}{2\bp\bq} = \xi\frac{Q^2 - Q'^2 + t/2}{Q^2+Q'^2} \,,
\end{gather}
as well as the ``average'' virtuality:
\begin{equation}
    \bQ^2 = -\bq^2 = \frac{1}{2}\left(Q^2 - Q'^2 + \frac{t}{2}\right)\,.
\end{equation}
It can be shown that $\xi\in (0, 1]$, while $\bxB$ and $\bQ^2$ can be either positive or negative depending on the relative magnitude between the spacelike ($Q^2$) and timelike ($Q'^2$) virtualities:
\begin{equation}
    \bxB = \xi\frac{2\bQ^2}{Q^2+Q'^2}\,.
\end{equation}
For completeness we also specify the range of Mandelstam variable $t$  allowed by the kinematics of the process:
\begin{align}
    t_{0}(t_{1}) = & -\frac{1}{4x_B(1-x_B) + \eps^2}\Bigg\{ 
    2[(1-x_B)Q^2 - x_BQ'^2] + \eps^2(Q^2-Q'^2) \nonumber\\
    & \mp 2\sqrt{1+\eps^2}\sqrt{[(1-x_B)Q^2 - x_BQ'^2]^2 - \eps^2Q^2Q'^2}
    \Bigg\} \,, \label{t0t1def}
\end{align}
where $t_{0}(t_{1})$ corresponds to $-(+)$ sign and minimal (maximal) absolute value of $t$.

\subsection{Relations to Trento and BDP frames}

Throughout the text we consequently work with angles defined in TRF-I and boosted TRF-II frames. In phenomenological applications it is however desired to use Trento frame \cite{trento} instead of TRF-I, and BDP frame \cite{Berger:2001xd} instead of TRF-II, as they are used in modern phenomenology and measurements of DVCS and TCS processes. Here, we give a concise prescription how to translate angles between the frames:
\begin{equation}
    \phi = \begin{dcases}
       \pi - \phiTrento, & \mbox{if }\phiTrento\in [0, \pi] \\
       3\pi - \phiTrento, & \mbox{if }\phiTrento\in(\pi, 2\pi)
    \end{dcases}\,,
\end{equation}
\begin{equation}
    \varphi_{S} = \begin{dcases}
    \pi - \phiSTrento, & \mbox{if }\phiSTrento\in [0, \pi] \\
       3\pi - \phiSTrento, & \mbox{if }\phiSTrento\in(\pi, 2\pi)
    \end{dcases}\,,
\end{equation}
and 
\begin{align}
    \sin\thetaLBDP & = \sqrt{(\ci\sin\thetaL\cos\phiL + \si\cos\thetaL)^2 + \sin^2\thetaL\sin^2\phiL} \,,\\
    \sin\phiLBDP & = \sin\thetaL\sin\phiL/\sin\thetaLBDP \,,\\
    \cos\phiLBDP & = (\ci\sin\thetaL\cos\phiL + \si\cos\thetaL)/\sin\thetaLBDP \,,
\end{align}
where $\ci = \cos\chi$ and $\si = \sin\chi$. The angle $\chi$ accounts for the rotation between TRF-II and BDP frames so that:
\begin{align}
    \ci & = \frac{p'^0\sinh\zeta - p'^3\cosh\zeta}{\sqrt{(p'^1)^2 + (p'^0\sinh\zeta - p'^3\cosh\zeta)^2}} \,,\\
    \si & = \frac{p'^1}{\sqrt{(p'^1)^2 + (p'^0\sinh\zeta - p'^3\cosh\zeta)^2}} \,,
\end{align}
where $\zeta = \arctanh{\modv}$ is the rapidity for the boost from TRF-II to the muon-antimuon CM frame. The components of $p'$ can be computed with momenta in Eqs.~(\ref{q'-TRFII})-(\ref{p-TRFII}) as $p' = p + q - q'$.
\section{Amplitudes and cross-section}
\label{sectAmp}

Electroproduction of a massless lepton pair can be described in terms of seven variables, so that the differential cross-section reads \cite{Belitsky:2002tf, guidal2003}:
\begin{equation}\label{xsec}
    \frac{d^7\sigma}{dx_B dQ^2 dQ'^2 d|t| d\phi d\Omega_\ell} = \frac{\alpha_{\rm em}^4}{16(2\pi)^3} \frac{x_By^2}{Q^4\sqrt{1+\eps^2}}\left|\frac{\M}{e^4}\right|^2 \,,
\end{equation}
where $\M$ stands for the amplitude of the process. If the target is polarized transversely, cross-section becomes dependent also on $\vphiS$ angle defined in Fig.~\ref{figure::DDVCSsetUp}. In such case, one should consider an 8-fold differential cross-section: $d^8\sigma/(dx_B dQ^2 dQ'^2 d|t| d\phi d\Omega_\ell d\vphiS) = d^7\sigma/(dx_B dQ^2 dQ'^2 d|t| d\phi d\Omega_\ell)\times (2\pi d\vphiS)^{-1}$.

The amplitude $\M$ receives contributions from all sub-processes (and their crossed partners) depicted in Fig.~\ref{figure::RelevantDiagrams}:
\begin{equation}
    i\M = i\M_{\rm DDVCS} + i\M_{\rm BH1} + i\M_{\rm BH1X} + i\M_{\rm BH2} + i\M_{\rm BH2X} \,.
\end{equation}
In the following we will separately evaluate each sub-amplitude. For this purpose we will make use of techniques developed by Kleiss and Stirling (KS) in Refs.~\cite{Kleiss:1984dp, Kleiss:1985yh}.

\subsection{DDVCS amplitude}

\begin{figure}[ht!]
    \centering
    \includegraphics{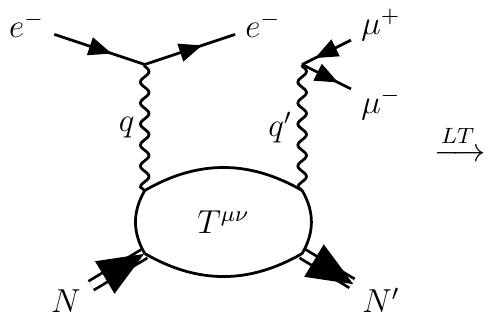}
    \includegraphics{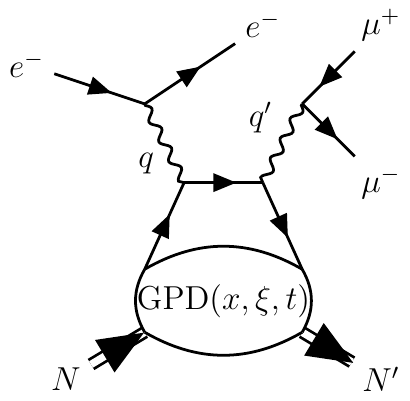}
    \caption{General DDVCS diagram in terms of Compton tensor $T^{\mu\nu}$ and its leading-twist (LT) approximation. The crossed diagram is not shown.} 
    \label{figure::ddvcsGeneral->LT}
\end{figure}

The amplitude of DDVCS contribution is the only one related to the internal distribution of partons in the hadron. It reads:
\begin{equation}
    i\M_{\rm DDVCS} = \frac{ie^4 \bar{u}(\ell_-,s_\ell)\gamma_\mu v(\ell_+,s_\ell)\bar{u}(k',s)\gamma_\nu u(k,s) }{(q^2+i0)(q'^2+i0)} T^{\mu\nu}_{s_2s_1}\,.
\end{equation}
Here, the Compton tensor $T^{\mu\nu}_{s_2s_1}$ is given by:
\begin{equation}\label{comptonTensor}
    T_{s_2s_1}^{\mu\nu} = i\int d^4z\ e^{i\bq z}\langle p',s_2|\mathcal{T}\{j^\mu(z/2)j^\nu(-z/2)\}|p,s_1\rangle,
\end{equation}
where $j^\mu(x) = \sum_f \frac{e_f}{e}\bar{\q}_f(x)\gamma^\mu\q_f(x)$ accounts for quark currents with flavors $f$ and electric charges $e_f$, and where $s_2, s_1 = \pm$ stand for hadron's helicity in final and initial state, respectively. 

To determine the leading terms in the Bj{\"o}rken regime it is convenient to define two light-like vectors, $n$ and $n^\star$, where $nn^\star = 1$. These light-like vectors are combinations of $\bp$ and $\bq$: 
\begin{align}
    n^\mu & = \frac{1}{\bp\bq R}\bq^\mu - \frac{1 - R}{ 2\bp\bq\bxB\delta^2 R}\bp^\mu \,,\\
    n^{\star\mu} & = -\frac{\bxB\delta^2}{R}\bq^\mu + \frac{1+R}{2R}\bp^\mu \,,
\end{align}
where $R = \sqrt{1+4\bxB^2\delta^2}$ and $\delta^2 = (M^2 - t/4)/(2\bp\bq\bxB)$. They allow for the decomposition of any four-vector in the following way: $v^\mu=v^-n^\mu + v^+n^{\star\mu} + v_\perp^\mu$. 

The Compton tensor introduced in Eq.~\eqref{comptonTensor} was computed to twist-3 accuracy in Ref.~\cite{Belitsky_2000}. For the purposes of our work it is enough to stay at leading twist (LT) and LO in $\alpha_s$, which corresponds to the  following decomposition:
\begin{equation}\label{comptonTensor_spinors}
    T^{\mu\nu}_{s_2s_2} = T^{(V)\mu\nu}\bar{u}(p',s_2)\left[ (\cffh + \cffe)\slashed{n} - \frac{\cffe}{M}\bp^+ \right] u(p, s_1) + T^{(A)\mu\nu}\bar{u}(p',s_2)\left[ \cffht\slashed{n} + \frac{\cffet}{2M}\Delta^+ \right]\gamma^5u(p,s_1)\,.
\end{equation}
Here, Compton form factors (CFFs) are defined as:
\begin{align}
    (\cffh,\cffe)(\bxB, \xi, t) & = \sum_{f=\{u, d, s\}}\int_{-1}^1 dx\ C_f^{(-)}(x, \bxB)(H_f, E_f)(x, \xi, t) \,, \label{cffHE} \\
    (\cffht,\cffet)(\bxB, \xi, t) & = \sum_{f=\{u, d, s\}}\int_{-1}^1 dx\ C_f^{(+)}(x, \bxB)(\widetilde{H}_f, \widetilde{E}_f)(x, \xi, t) \,, \label{cffHtEt}
\end{align}
where $C^{(\pm)}_f$ are hard scattering coefficient functions, which at LO read\footnote{This form of the LO coefficient function reflects the fact that, as for DVCS and TCS, only the charge conjugation even part of GPDs contribute to the DDVCS process. To access the complementary charge conjugation odd part of quark GPDs, one needs to study other processes like diphoton photo- or electroproduction \cite{Pedrak:2017cpp, Pedrak:2020mfm, Grocholski:2021man, Grocholski:2022rqj}. To access the chiral-odd quark GPDs, processes containing mesons in the final state are necessary \cite{Ivanov:2002jj, Boussarie:2016qop, Duplancic:2023kwe}.
}
\begin{equation}\label{C+-}
    C^{(\pm)}_f(x, \bxB) = \left(\frac{e_f}{e}\right)^2\left( \frac{1}{\bxB - x - i0} \pm \frac{1}{\bxB + x - i0} \right)\,,
\end{equation}
and where $H, E, \widetilde{H}$ and $\widetilde{E}$ are GPDs related to quark correlators (notice that $\bp^+ = 1$) as
%($n$ is a light-like vector fixing the ``plus'' direction)
\begin{align}
    \int \frac{d\lambda}{2\pi}\ e^{-i\lambda x}\langle p',s_2| \bar{\q}_f(\lambda n/2)\slashed{n}\q_f(-\lambda n/2) |p, s_1\rangle & =\bar{u}(p',s_2)\left[ (\cffh + \cffe)\slashed{n} - \frac{\cffe}{M}\bp^+ \right] u(p, s_1) \equiv \mathcal{J}^+_{s_2s_1} \label{Jcal}\,,\\
    \int \frac{d\lambda}{2\pi}\ e^{-i\lambda x}\langle p',s_2| \bar{\q}_f(\lambda n/2)\slashed{n}\gamma^5\q_f(-\lambda n/2) |p, s_1\rangle & = \bar{u}(p',s_2)\left[ \cffht\slashed{n} + \frac{\cffet}{2M}\Delta^+ \right]\gamma^5u(p,s_1) \equiv \mathcal{J}_{s_2s_1}^{(5)+}\,. \label{J5cal}
\end{align}
Finally, the Lorentz components have been isolated and concealed in the following tensors: 
\begin{align}
    T^{(V)\mu\nu} & = -\frac{1}{2}\left( g^{\mu\nu} - \frac{q^\mu q'^\nu}{qq'} \right) + \frac{\bxB}{\bp\bq}\left( \bp^\mu - \frac{\bp q'}{qq'}q^\mu \right)\left( \bp^\nu - \frac{\bp q}{qq'}q'^\nu \right) \label{T1}\,, \\
    T^{(A)\mu\nu} & = \frac{i}{2\bp\bq}\epsilon_{\theta\lambda\rho\sigma}\bp^\rho\bq^\sigma \left( g^{\mu\theta} - \frac{\bp^\mu q'^\theta}{\bp q'} \right) \left( g^{\nu\lambda} - \frac{\bp^\nu q^\lambda}{\bp q} \right)\,. \label{T2}
\end{align}

To LT, the dominant terms in the amplitude arise from keeping the following structures in tensors above:
\begin{align}
    T^{(V)\mu\nu} & = -\frac{1}{2}(g^{\mu\nu} - n^\mu n^{\star\nu} - n^\nu n^{\star\mu}) \equiv -\frac{1}{2}g_\perp^{\mu\nu}\,,\\
    T^{(A)\mu\nu} & = -\frac{i}{2}{\epsilon^{\mu\nu}}_{\rho\sigma} n^\rho n^{\star\sigma} \equiv -\frac{i}{2}\epsilon^{\mu\nu}_\perp\,,
\end{align}
which are used to define the vector and axial components of the DDVCS amplitude:
\begin{equation}
    i\M_{\rm DDVCS} = \frac{-ie^4}{(Q^2-i0)(Q'^2+i0)}\left( i\M^{(V)}_{\rm DDVCS} + i\M^{(A)}_{\rm DDVCS} \right)\,.
\end{equation}
Here, the first term (the vector contribution) corresponds to $T^{(V)}$, while the second one (the axial contribution) to $T^{(A)}$. In what follows we study these two contributions separately.

\subsubsection{Vector contribution to the DDVCS amplitude}

Up to photon propagators and factors $ie^4$, the vector amplitude may be written as:
\begin{equation}
    i\M_{\rm DDVCS} ^{(V)} = -\frac{g_\perp^{\mu\nu}}{2}\bar{u}(\ell_-,s_\ell)\gamma_\mu v(\ell_+,s_\ell)\bar{u}(k',s)\gamma_\nu u(k,s) \mathcal{J}^+_{s_2s_1}\,,
\end{equation}
where $s_\ell, s = \pm$ stand for muon's and electron's helicities, respectively. $\mathcal{J}^+$ is given by Eq.~\eqref{Jcal} and can be further decomposed into:
\begin{equation}\label{JcalDecomposition}
    \mathcal{J}^+_{s_2s_1} = (\cffh + \cffe) \Joplus_{s_2s_1} - \frac{\cffe}{M}\Jt_{s_2s_1}\,,
\end{equation}
where
\begin{equation}\label{Jcal1muJcal2}
    \mathcal{J}^{(1)\mu}_{s_2s_1} =\bar{u}(p',s_2)\gamma^\mu u(p, s_1)
    , \quad
    \mathcal{J}^{(2)}_{s_2s_1} =\bar{u}(p',s_2) u(p, s_1)\,.
\end{equation}
To use the KS methods for massive spinors, hadron momenta $p$ and $p'$ have to be decomposed by means of auxiliary light-like vectors, this is, $p = r_1+r_2$ and $p' = r'_1+r'_2$. The decomposition of nucleon spinors into massles ones reads:
\begin{align}
    u(p,+) & = \frac{s(r_1, r_2)}{M}u(r_1,+) + u(r_2,-) \,,\label{u+massive} \\
    u(p,-) & = \frac{t(r_1, r_2)}{m}u(r_1,-) + u(r_2,+) \,,\label{u-massive} 
\end{align}
where for two light-like vector $a$ and $b$:
\begin{align}
    s(a, b) & = \bar{u}(a,+)u(b,-) = -s(b, a)\,, \label{sKS_def}\\
    t(a, b) & = \bar{u}(a,-)u(b,+) = [s(b, a)]^*\,. \label{tKS_def}
\end{align}
Explicit computation of bilinears above shows that $s(a, b)$ acquires the simple form (see Eq.~(3.1) in Ref.~\cite{Kleiss:1985yh}):
\begin{equation}\label{sKS_expression}
    s(a, b) = (a^2 + ia^3)\sqrt{\frac{b^0 - b^1}{a^0 - a^1}} - (a\leftrightarrow b)\,,
\end{equation}
as long as $a\cdot\kappa_0\neq 0$ and $b\cdot\kappa_0 \neq 0$ with $\kappa^\mu_0 = (1, 1, 0, 0)$.

By making use of the above formulae we get the following expressions:
\begin{equation}\label{Jcal1mu}
    \mathcal{J}^{(1)\mu}_{s_2s_1} = Y_{s_2s_1}\bar{u}(r'_{s_2},+)\gamma^\mu u(r_{s_1},+) + Z_{s_2s_1}\bar{u}(r'_{-s_2},-)\gamma^\mu u(r_{-s_1},-)
\end{equation}
where $r'_{s_2} = r'_1\delta_{s_2+} + r'_2\delta_{s_2-}$ and $r_{s_1} = r_1\delta_{s_1+} + r_2\delta_{s_1-}$. Phases\begin{NoHyper}\footnote{$Y$ and $Z$ have unit modulus as $|s(r_1, r_2)|^2 = 2r_1r_2 = M^2$. Likewise, for $s\leftrightarrow t$ and/or $r_{1, 2}\leftrightarrow r'_{1, 2}$.}\end{NoHyper} $Y, Z$ read:
\begin{equation}
    Y_{s_2s_1} = \delta_{s_2+}\delta_{s_1+}\frac{t(r'_2,r'_1)s(r_1,r_2)}{M^2} + \delta_{s_2+}\delta_{s_1-}\frac{t(r'_2,r'_1)}{M} + \delta_{s_2-}\delta_{s_1+}\frac{s(r_1,r_2)}{M} + \delta_{s_2-}\delta_{s_1-}
\end{equation}
and
\begin{equation}
    Z_{s_2s_1} =
    \delta_{s_2-}\delta_{s_1-}\frac{s(r'_2,r'_1)t(r_1,r_2)}{M^2} + \delta_{s_2-}\delta_{s_1+}\frac{s(r'_2,r'_1)}{M} + \delta_{s_2+}\delta_{s_1-}\frac{t(r_1,r_2)}{M} + \delta_{s_2+}\delta_{s_1+}\,.
\end{equation}
A similar calculation for $\Jt_{s_2s_1}$ yields:
\begin{align}\label{Jcal2}
    \Jt_{s_2s_1} = &\phantom{+} \delta_{s_2+}\delta_{s_1+}
    \left[
    \frac{t(r'_2,r'_1)s(r'_1,r_2)}{M} + \frac{t(r'_2,r_1)s(r_1,r_2)}{M} 
    \right]
    \nonumber\\
    & + \delta_{s_2+}\delta_{s_1-}\left[ \frac{t(r'_2,r'_1)t(r_1,r_2)s(r'_1,r_1)}{M^2} + t(r'_2,r_2) \right] \nonumber\\
    & + \delta_{s_2-}\delta_{s_1+}\left[ \frac{s(r'_2,r'_1)s(r_1,r_2)t(r'_1,r_1)}{M^2} + s(r'_2,r_2) \right] \nonumber\\
    & +  \delta_{s_2-}\delta_{s_1-}
    \left[
    \frac{s(r'_2,r'_1)t(r'_1,r_2)}{M} + \frac{s(r'_2,r_1)t(r_1,r_2)}{M}
    \right]\,.
\end{align}
For further calculations it is useful to define two other scalars, namely the contraction of two currents: 
\begin{align}\label{function_f}
    f(\lambda, k_0, k_1; \lambda', k_2, k_3) = & \bar{u}(k_0,\lambda)\gamma^\mu u(k_1, \lambda)\bar{u}(k_2,\lambda')\gamma_\mu u(k_3,\lambda') \nonumber\\
    = & 2 [ s(k_2,k_1)t(k_0,k_3)\delta_{\lambda-}\delta_{\lambda'+} + t(k_2,k_1)s(k_0,k_3)\delta_{\lambda+}\delta_{\lambda'-} \nonumber\\
    & + s(k_2,k_0)t(k_1,k_3)\delta_{\lambda+}\delta_{\lambda'+} + t(k_2,k_0)s(k_1,k_3)\delta_{\lambda-}\delta_{\lambda'-} ]\,,
\end{align}
and the contraction of a current with a light-like vector $a$:
\begin{align}\label{function_g}
    g(s, \ell, a, k) = & \bar{u}(\ell, s)\slashed{a}u(k, s) \nonumber\\
    = & \delta_{s+}s(\ell,a)t(a,k) + \delta_{s-}t(\ell,a)s(a,k)\,.
\end{align}

By contracting Eq.~\eqref{Jcal1mu} with vector $n$ we arrive to:
\begin{equation}
    \Joplus_{s_2s_1}= \mathcal{J}^{(1)\mu}_{s_2s_1}n_\mu = Y_{s_2s_1}g(+,r'_{s_2},n,r_{s_1}) + Z_{s_2s_1}g(-,r'_{-s_2},n,r_{-s_1}) \,.
\end{equation}

Finally, the vector contribution to the DDVCS amplitude is:
\begin{align}\label{iM1ddvcs_final}
    i\M^{(V)}_{\rm DDVCS} = & -\frac{1}{2}\Bigg[ f(s_\ell, \ell_-, \ell_+; s, k', k) -  g(s_\ell,\ell_-,n^\star,\ell_+)g(s, k',n,k) - g(s_\ell,\ell_-,n,\ell_+)g(s, k',n^\star,k) \Bigg] \nonumber\\
    & \times \Bigg[ (\cffh + \cffe) [ Y_{s_2s_1}g(+,r'_{s_2},n,r_{s_1}) + Z_{s_2s_1}g(-,r'_{-s_2},n,r_{-s_1}) ] - \frac{\cffe}{M} \Jt_{s_2s_1} \Bigg]\,.
\end{align}
\subsubsection{Axial contribution to the DDVCS amplitude}
The axial amplitude $T^{(A)}$  may be written up to photon propagators and factors $ie^4$ as:
\begin{equation}\label{iM2ddvcs_initial}
    i\M^{(A)}_{\rm DDVCS} = \frac{-i\epsilon_\perp^{\mu\nu}}{2}\bar{u}(\ell_-,s_\ell)\gamma_\mu v(\ell_+,s_\ell)\bar{u}(k',s)\gamma_\nu u(k, s)\mathcal{J}^{(5)+}_{s_2s_1}\,,
\end{equation}
where $\mathcal{J}^{(5)+}$ was defined in Eq.~\eqref{J5cal}. We distinguish $\Jofplus$ and $\Jtfplus$:
\begin{equation}
    \mathcal{J}^{(5)+}_{s_2s_1} = \cffht\Jofplus_{s_2s_1} + \cffet\frac{\Delta^+}{2M}\Jtfplus_{s_2s_1}\,,
\end{equation}
such that
\begin{align}
    \Jofplus_{s_2s_1} & = \bar{u}(p',s_2)\slashed{n}\gamma^5u(p,s_1) \label{J15+}\,,\\
    \Jtfplus_{s_2s_1} & = \bar{u}(p',s_2)\gamma^5u(p,s_1) \,.\label{J25}
\end{align}

Similarly to the vector case, we can express the currents in terms of the scalar functions:
\begin{align}
    \Jofplus_{s_2s_1} = & 
    \phantom {-} \delta_{s_2+}\delta_{s_1+}\Bigg[ \frac{t(r'_2,r'_1)s(r_1,r_2)t(n,r_1)s(r'_1,n)}{M^2} - s(n,r_2)t(r'_2,n) \Bigg] 
    \nonumber \\  & 
    - \delta_{s_2-}\delta_{s_1-}
    \Bigg[ \frac{s(r'_2,r'_1)t(r_1,r_2)s(n,r_1)t(r'_1,n)}{M^2} - t(n,r_2)s(r'_2,n) \Bigg] 
    \nonumber \\  &
    + \delta_{s_2+}\delta_{s_1-} \frac{t(r'_2,r'_1)t(n,r_2)s(r'_1,n) - t(r_1,r_2)s(n,r_1)t(r'_2,n)}{M} 
    \nonumber \\  &
    - \delta_{s_2-}\delta_{s_1+}
    \frac{s(r'_2,r'_1)s(n,r_2)t(r'_1,n) - s(r_1,r_2)t(n,r_1)s(r'_2,n)}{M} 
\end{align}
and
\begin{align}
    \Jtfplus_{s_2s_1} =  & \phantom{-} \delta_{s_2+}\delta_{s_1+}\frac{ s(r_1,r_2)t(r'_2,r_1) - t(r'_2,r'_1)s(r'_1,r_2) }{M} 
     + \delta_{s_2+}\delta_{s_1-}\Bigg[ t(r'_2,r_2) - \frac{t(r'_2,r'_1)t(r_1,r_2)s(r'_1,r_1)}{M^2} \Bigg] \nonumber\\
    & - \delta_{s_2-}\delta_{s_1-}\frac{ t(r_1,r_2)s(r'_2,r_1) - s(r'_2,r'_1)t(r'_1,r_2) }{M} 
     - \delta_{s_2-}\delta_{s_1+} \Bigg[ s(r'_2,r_2) - \frac{s(r'_2,r'_1)s(r_1,r_2)t(r'_1,r_1)}{M^2} \Bigg]\,.
\end{align}

Because of $\epsilon_\perp$-structure in Eq.~\eqref{iM2ddvcs_initial} we cannot contract the above lepton currents, hence we need to compute them explicitly. The massless lepton current for muon-antimuon pair can be written as:
\begin{align}
    j_\mu(s_\ell, \ell_-, \ell_+) = & \bar{u}(\ell_-,s_\ell)\gamma_\mu v(\ell_+,s_\ell) \nonumber\\
    = & \frac{1}{2N_{\ell_-\ell_+}}{\rm tr}\left\{ \slashed{\ell}_-\gamma_\mu\slashed{\ell}_+\omega_{-s_\ell}\slashed{\K}_0 \right\} \nonumber\\
    = & \frac{1}{N_{\ell_-\ell_+}} \{ \ell_{-,\mu}(\ell_+\K_0) + \ell_{+,\mu}(\ell_-\K_0) - \K_{0,\mu}Q'^2/2 + is_\ell\epsilon_{\mu\alpha\beta\gamma}\ell_-^\alpha\ell_+^\beta\K_0^\gamma
    \}\,,
\end{align}
where $N_{\ell_-\ell_+} = \sqrt{(\ell_-\K_0)(\ell_+\K_0)}$.
Likewise, for the electron current ($N_{k'k} = \sqrt{(k'\K_0)(k\K_0)}$):
\begin{equation}
    j_\mu(s, k', k) = \frac{1}{N_{k'k}} \{ k'_{\mu}(k\K_0) + k_{\mu}(k'\K_0) - \K_{0,\mu}Q^2/2 + is\epsilon_{\mu\alpha\beta\gamma}k'^\alpha k^\beta\K_0^\gamma
    \}\,.
\end{equation}

Finally, the axial contribution is given by:
\begin{equation}\label{iM2ddvcs_final}
    i\M^{(A)}_{\rm DDVCS} = \frac{-i}{2} \epsilon^{\mu\nu}_\perp j_\mu(s_\ell,\ell_-,\ell_+)j_\nu(s, k', k)\left[ \cffht\Jofplus_{s_2s_1} + \cffet\frac{\Delta^+}{2M}\Jtfplus_{s_2s_1} \right]\,.
\end{equation}

Violation of gauge invariance due to a truncation of the twist expansion of the Compton tensor is a problem discussed in papers like \cite{Anikin_2000, Vanderhaeghen_1999} and more recently in \cite{Braun:2012hq, Braun:2014sta}. If any of the photon momenta carries a perpendicular component (as happens in TRF-II described in Sect.~\ref{sec::frames}), gauge invariance is violated by terms of order $O(\Delta_\perp/\sqrt{2\bp\bq})$, which can be considered twist-3 effects. In Ref.~\cite{Belitsky_2000} it was proven that by going to twist-3, violation is of order twist-4 and so on. Despite restoring gauge invariance is possible twist-by-twist, the existence of gauge symmetry-breaking terms affects predictions at LT. There are two ways to deal with it: 1) Lorentz transformation to a reference frame where photons do not carry perpendicular components, or 2) evaluation of hard part, which includes the Compton tensor's Lorentz structures, at $t = t_0$ (defined by Eq.~\eqref{t0t1def}). Choosing to evaluate at $t_0$ ensures that higher-twist corrections proportional to $\Delta_\perp^\mu$ vanish, avoiding violation of gauge invariance. This evaluation is consistent with the longitudinal factorization that is at the core of GPD description. In our phenomenological study in Sect.~\ref{sectObs}, we opt for this second option. As a consequence, the vector $n$, which defines the ``plus'' direction is also evaluated at $t = t_0$.

\subsection{The first Bethe-Heitler amplitude}

\begin{figure}[!ht]
    \centering
    \includegraphics{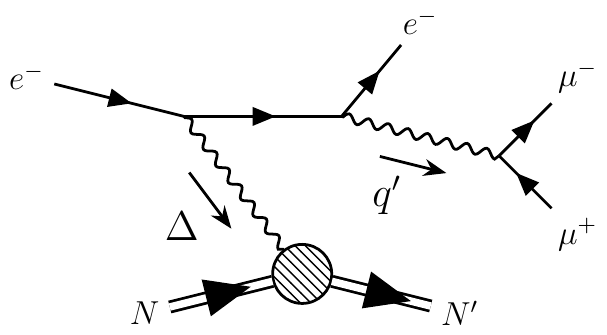}
    \includegraphics{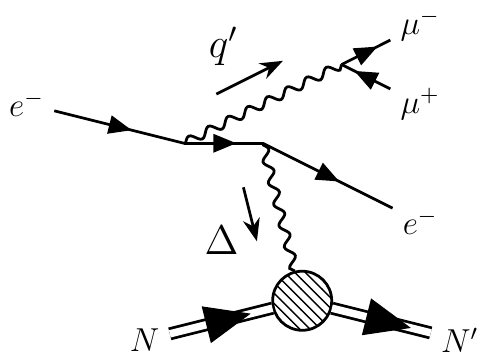}
    \caption{Diagrams for Bethe-Heitler 1 (BH1) and its crossed partner (BH1X) for electroproduction of muon pairs.}
    \label{figure::BH1}
\end{figure}

Now we describe the amplitude of BH1 and its crossed partner BH1X, both depicted in Fig.~\ref{figure::BH1}.

The amplitude of BH1 reads:
\begin{equation}
    i\M_{\rm BH1} = \frac{ie^4 \bar{u}(\ell_-, s_\ell)\gamma^\beta v(\ell_+,s_\ell)\bar{u}(k',s)\gamma_\beta(\slashed{k}-\slashed{\Delta})\gamma_\alpha u(k, s) J^\alpha_{s_2s_1} }{(Q'^2+i0)(t+i0)((k-\Delta)^2+i0)}\,,
\end{equation}
where the electromagnetic hadronic current is defined by means of Dirac, $F_1(t)$, and Pauli, $F_2(t)$, elastic form factors (FFs):
\begin{align}
    J^\alpha_{s_2s_1} =~ & \langle p',s_2| \sum_f \frac{e_f}{e}\bar{\q}_f(0)\gamma^\alpha \q_f(0) |p,s_1\rangle 
    %\ \label{Jalpha_def} \\
    =~  \bar{u}(p',s_2)\left[ (F_1+F_2)\gamma^\alpha - \frac{F_2}{M}\bp^\alpha \right]u(p,s_1)\,. \label{Jalpha_spinors}
\end{align}

The structure of this current is the same as that of $\mathcal{J}^+$, see Eq.~\eqref{Jcal}. Therefore, up to CFF $\leftrightarrow$ FF replacement, one may apply the decomposition \eqref{JcalDecomposition}:
\begin{equation}\label{Jalpha}
    J^\alpha_{s_2s_1} = (F_1 + F_2) \mathcal{J}^{(1)\alpha}_{s_2s_1} - \frac{F_2}{M}\bp^\alpha\Jt_{s_2s_1}\,,
\end{equation}
where $\mathcal{J}^{(1)\alpha}$ and $\Jt$ are defined in Eqs.~\eqref{Jcal1mu} and \eqref{Jcal2}, respectively. 

With the decomposition \eqref{Jalpha} the evaluation of $i\M_{\rm BH1}$ can be separated into two parts related to $\mathcal{J}^{(1)}$ and $\Jt$:
\begin{equation}\label{iMbh1_splitting}
    i\M_{\rm BH1} = \frac{ie^4 \left( i\M^{(1)}_{\rm BH1} + i\M^{(2)}_{\rm BH1} \right) }{(Q'^2+i0)(t+i0)((k-\Delta)^2+i0)}\,.
\end{equation}

The first term in the numerator of Eq.~\eqref{iMbh1_splitting}, by means of the scalar function $f$ introduced in Eq.~\eqref{function_f}, can be expressed as:
\begin{equation}\label{eq:iM1BH1}
    i\M^{(1)}_{\rm BH1} = (F_1+F_2)\sum_Lf(s_\ell,\ell_-,\ell_+; s, k', L)\Big( Y_{s_2s_1}f(s,L,k; +,r'_{s_2}, r_{s_1}) + Z_{s_2s_1}f(s,L,k; -,r'_{-s_2}, r_{-s_1}) \Big)\,,
\end{equation}
where $L\in\{k',\ell_-,\ell_+\}$.

The second term in Eq.~(\ref{iMbh1_splitting}), after expanding $\bp$ in the sum of light-like vectors $R\in\{r_1, r_2, r'_1, r'_2\}$, has the following form:
\begin{equation}\label{eq:iM2BH1}
    i\M^{(2)}_{\rm BH1} = -\frac{F_2}{2M}\Jt_{s_2s_1}\sum_{L, R}f(s_\ell,\ell_-,\ell_+; s,k',L)g(s, L, R, k)\,.
\end{equation}

The amplitude of crossed BH1 reads:
\begin{equation}
    i\M_{\rm BH1X} = \frac{ie^4 \bar{u}(\ell_-,s_\ell)\gamma^\beta v(\ell_+,s_\ell)\bar{u}(k',s)\gamma_\alpha(\slashed{k}' + \slashed{\Delta})\gamma_\beta u(k,s)J^\alpha_{s_2s_1} }{(q'^2+i0)(t+i0)((k'+\Delta)^2+i0)}\,.
\end{equation}

Analogously to Eqs.~\eqref{iMbh1_splitting}, \eqref{eq:iM1BH1} and \eqref{eq:iM2BH1}:
\begin{equation}
    i\M_{\rm BH1X} = \frac{ie^4 \left( i\M^{(1)}_{\rm BH1X} + i\M^{(2)}_{\rm BH1X} \right) }{(Q'^2+i0)(t+i0)((k'+\Delta)^2+i0)}\,,
\end{equation}
and
\begin{align}
    i\M^{(1)}_{\rm BH1X} = & (F_1+F_2)\sum_L\sigma(L) f(s_\ell,\ell_-,\ell_+; s, L, k') \Big( Y_{s_2s_1}f(s, k', L; +, r'_{s_2},r_{s_1}) + Z_{s_2s_1}f(s, k', L;-,r'_{-s_2},r_{-s_1}) \Big) \,,\\
    i\M^{(2)}_{\rm BH1X} = & -\frac{F_2}{2M}\Jt_{s_2s_1}\sum_{L,R}\sigma(L)f(s_\ell,\ell_-,\ell_+; s, L, k)g(s,k',R,L)\,,
\end{align}
where $L\in\{k,\ell_-,\ell_+\}$, $R\in\{r_1,r_2,r'_1,r'_2\}$ and $\sigma(k) = +1$, $\sigma(\ell_-)= \sigma(\ell_+)=-1$.

\subsection{The second Bethe-Heitler amplitude}

\begin{figure}[!ht]
    \centering
    \includegraphics{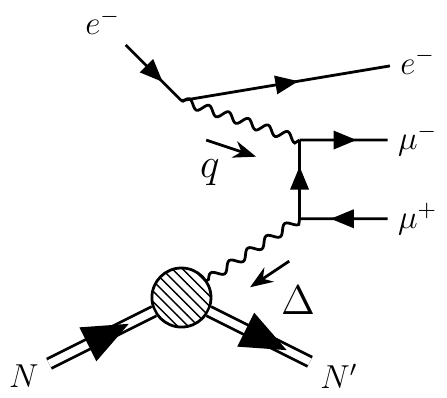}
    \includegraphics{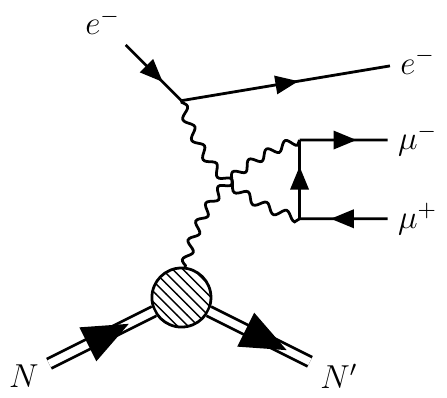}
    \caption{Diagrams for Bethe-Heitler 2 (BH2) and its crossed partner (BH2X) for electroproduction of muon pairs.}
    \label{figure::BH2}
\end{figure}

BH2 and its crossed partner are shown in Fig.~\ref{figure::BH2}. The amplitude of the former reads: 
\begin{equation}
    i\M_{\rm BH2} = \frac{ie^4 \bar{u}(k',s)\gamma^\beta u(k,s)\bar{u}(\ell_-,s_\ell)\gamma_\beta(\slashed{k} - \slashed{k}' - \slashed{\ell}_-)\gamma_\alpha v(\ell_+,s_\ell)J^\alpha_{s_2s_1} }{(Q^2-i0)(t+i0)((q-\ell_-)^2+i0)}\,,
\end{equation}
and, as for BH1, can be split into two terms corresponding to elements of Eq.~\eqref{Jalpha} as:
\begin{equation}
    i\M_{\rm BH2} = \frac{ie^4 \left( i\M^{(1)}_{\rm BH2} + i\M^{(2)}_{\rm BH2} \right) }{(Q^2-i0)(t+i0)((q-\ell_-)^2+i0)}\,.
\end{equation}

The same steps as presented in the previous sections lead to:
\begin{align}
    i\M^{(1)}_{\rm BH2} = & (F_1+F_2)\sum_L\sigma(L) f(s_\ell,\ell_-,L;s,k',k)\Big( Y_{s_2s_1}f(s_\ell,L,\ell_+;+,r'_{s_2},r_{s_1}) + Z_{s_2s_1}f(s_\ell,L,\ell_+;-,r'_{-s_2},r_{-s_1}) \Big) \,,\\
    i\M^{(2)}_{\rm BH2} = & \frac{-F_2}{2M}\Jt_{s_2s_1}\sum_{L, R}\sigma(L) f(s_\ell,\ell_-,L;s,k',k)g(s_\ell,L,R,\ell_+)\,,
\end{align}
where $L\in\{k,k',\ell_-\}$, $R\in\{r_1,r_2,r'_1,r'_2\}$, $\sigma(k) = +1$ and $\sigma(k') = \sigma(\ell_-) = -1$. 

The amplitude for the crossed partner of BH2 is given by:
\begin{equation}
    i\M_{\rm BH2X} = \frac{ie^4 \bar{u}(\ell_-,s_\ell)\gamma_\alpha(\slashed{k}- \slashed{k}' - \slashed{\ell_+})\gamma_\beta v(\ell_+,s_\ell)\bar{u}(k',s)\gamma^\beta u(k,s)J^\alpha_{s_2s_1} }{(Q'^2+i0)(t+i0)((q-\ell_+)^2 + i0)}\,.
\end{equation}
It can be expressed by:
\begin{equation}
    i\M_{\rm BH2X} =  \frac{-ie^4 \left( i\M^{(1)}_{\rm BH2X} + i\M^{(2)}_{\rm BH2X} \right) }{(Q^2-i0)(t+i0)((q-\ell_+)^2 + i0)}\,,
\end{equation}
for which
\begin{align}
    i\M^{(1)}_{\rm BH2X} = & (F_1+F_2)\sum_L\sigma(L)f(s_\ell,L,\ell_+;s, k',k)\Big( Y_{s_2s_1}f(s_\ell, \ell_-,L;+,r'_{s_2},r_{s_1}) + Z_{s_2s_1}f(s_\ell,\ell_-,L;-,r'_{-s_2},r_{-s_1}) \Big) \,,\\
    i\M^{(2)}_{\rm BH2X} = & -\frac{F_2}{2M}\Jt_{s_2s_1}\sum_{L,R}\sigma(L) f(s_\ell,L,\ell_+;s,k',k)g(s_\ell,\ell_-,R,L)\,,
\end{align}
with $L\in\{k,k',\ell_+\}$, $R\in\{r_1,r_2,r'_1,r'_2\}$ and $\sigma(k) = +1, \sigma(k')=\sigma(\ell_+)=-1$.

\subsection{Polarized target case}

Although we are not presenting the results for longitudinally and transversely polarized targets, for completeness we describe in the following how to address such cases within the Kleiss-Stirling approach. Thus far, hadron polarization denoted with index $s_1$ corresponds to the values $\pm$ for helicity with respect to the three-vector component of 
\begin{equation}
    s^\mu = (r_1^\mu - r_2^\mu)/M \,,
\end{equation}
with $M$ the target mass and $r_1,r_2$ two light-like vectors, such that hadron momentum can be written as $p = r_1+r_2$.

With respect to TRF-II axes, $s$ reads:
\begin{equation}
    s^\mu = (0, \hat{z}), \quad \hat{z} = (0, 0, 1)\,.
\end{equation}
Therefore, $\vec{s} = \hat{z}$ is parallel to the outgoing photon three-momentum $\vec{q'}$.
The relation between the quantization of helicity in direction $\hat{z}$ (denoted by $s_1=\pm$) and in another direction defined by three-vector\begin{NoHyper}\footnote{Angles $\phiS$ and $\thetaS$ are the azimuthal and polar orientations of $\vec{S}$ with respect to TRF-II.}\end{NoHyper} $\vec{S} = (\sin\thetaS\cos\phiS,\sin\thetaS\sin\phiS,\cos\thetaS)$ is:
\begin{align}
    | h_1 = + \rangle & = \cos(\thetaS/2)|s_1 = +\rangle + e^{i\phiS}\sin(\thetaS/2)|s_1 = -\rangle \,, \label{generalSpinStates+} \\
    | h_1 = - \rangle & = -e^{-i\phiS}\sin(\thetaS/2)|s_1 = +\rangle + \cos(\thetaS/2)|s_1 = -\rangle \,. \label{generalSpinStates-}
\end{align}

Introducing these spin states in the Compton tensor \eqref{comptonTensor} and the electromagnetic hadron current \eqref{Jalpha_spinors} is equivalent to relate spinors $u'(p,h_1)$ to $u(p,s_1)$ via:
\begin{equation}\label{generalSpinor}
     u'(p, h_1) = F_{h_1 +} u(p,+) + F_{h_1 -} u(p,-)\,,
\end{equation}
where matrix $F$ has been defined in accordance to \eqref{generalSpinStates+} and \eqref{generalSpinStates-} as
\begin{equation}
    F = \begin{pmatrix}
    \cos{\frac{\thetaS}{2}} & e^{i\phiS}\sin{\frac{\thetaS}{2}} \\
    -e^{-i\phiS}\sin{\frac{\thetaS}{2}} & \cos{\frac{\thetaS}{2}}
    \end{pmatrix} = \begin{pmatrix}
    F_{++} & F_{+-}\\
    F_{-+} & F_{--}
    \end{pmatrix}\,.
\end{equation}

Therefore, using (\ref{generalSpinor}) in Eqs.~(\ref{Jcal}), (\ref{J5cal}) and (\ref{Jalpha_spinors}), the correspondence between our current amplitudes, where target is polarized in direction $\hat{z}$ (index $s_1$), and the ones with a target polarized with respect to $\vec{S}$ (index $h_1$) reads:
\begin{equation}
    i\M(s_2,h_1) = F_{h_1+}i\M(s_2,s_1=+) + F_{h_1-}i\M(s_2,s_1=-)\,.
\end{equation}
We can orientate $\vec{S}$ in such a way that we define two types of target polarization: longitudinal ($\vec{S}\parallel\vec{k}$) and transverse  ($\vec{S}\perp\vec{k}$) to the electron beam.

\section{DVCS and TCS limits}\label{sectLimits}

In this section we numerically validate our results against DVCS and TCS limits, which were previously described in literature \cite{BELITSKY2014214,Berger:2001xd} and implemented in PARTONS framework. In these tests we utilize Goloskokov-Kroll GPD model (see for example \cite{Goloskokov_2007, Goloskokov_2007_2}), and the renormalization and factorization scales are $\mu_R^2 = \mu_F^2 = Q^2+Q'^2$. Furthermore, the skewness and generalized Bj{\"o}rken variables are evaluated at $t = 0$. The cross-sections are presented for Trento and BDP angles (see Sect. \ref{sectKin} for details).

The DVCS limit is obtained by taking $Q'^2 \to 0$, which for CFFs gives $\mathcal{F}(\rho, \xi, t)\xrightarrow{Q'^2\rightarrow 0}\mathcal{F}(\xi, \xi, t)$. To make DDVCS and DVCS cross-sections comparable, one must correct the former for the residual $Q'^2$ dependence and the splitting of the outgoing virtual-photon into the lepton pair:
\begin{equation}
    \int d\Omega_\ell \underbrace{\frac{d^7\sigma}{dx_B dQ^2 dQ^{\prime 2}d|t|d\phi d\Omega_\ell}}_{{\rm DDVCS}} \xrightarrow{Q^{\prime 2} \rightarrow 0} \underbrace{\left( \frac{d^4\sigma}{dx_B dQ^2 d|t|d\phi} \right)}_{{\rm DVCS}} \frac{\mathcal{N}}{Q^{\prime 2}} \,,
    \label{eq:limitPrescriptionDVCS}
\end{equation}
where $\mathcal{N} = \alpha_{\rm em}/(3\pi)$ \cite{Vanderhaeghen_1999}. 

The TCS limit, on the other hand, is obtained by taking $Q^2 \to 0$, which for Compton form factors gives $\mathcal{F}(\rho, \xi, t)\xrightarrow{Q^2\rightarrow 0}\mathcal{F}(-\xi, \xi, t)$. In term of cross-sections, we make the integration over $\phi$ angle and we include the photon flux, $\Gamma$, calculated within the equivalent photon approximation (EPA) \cite{kessler_epa, kessler_halArchives}: 
\begin{equation}
    \int d\phi \underbrace{\frac{d^7\sigma}{dx_B dQ^2 dQ^{\prime 2}d|t|d\phi d\Omega_\ell}}_{{\rm DDVCS}} \xrightarrow{Q^{2} \rightarrow 0} \underbrace{\left( \frac{d^4\sigma}{dQ^{\prime 2}d|t|d\Omega_\ell} \right)}_{{\rm TCS}} \frac{d^2\Gamma}{dx_{B}dQ^2} \,,
    \label{eq:limitPrescriptionTCS}
\end{equation}
where
\begin{equation}
    \frac{d^2\Gamma}{dx_{B}dQ^2} = \frac{\alpha_{\rm em}}{2\pi Q^2}\left( 1 + \frac{(1-y)^2}{y} - \frac{2(1-y)Q_{\rm min}^2}{yQ^2} \right)\frac{\nu}{Ex_B} \,.
\end{equation}
Here, 
\begin{equation}
    \nu = \frac{Q^2}{2 M x_B} 
\end{equation}
is the energy of the photon beam, while
\begin{equation}
    Q^2_{\rm min} = \frac{(y m_e)^2}{1-y}
\end{equation}
is the minimum value of the spacelike virtuality evaluated for the electron mass, $m_e$. We note that the prescriptions for both DVCS and TCS limits hold for each sub-process, i.e. BH, pure DDVCS and the interference. 

\begin{figure}[!ht]
    \centering
    \includegraphics[width=0.45\textwidth]{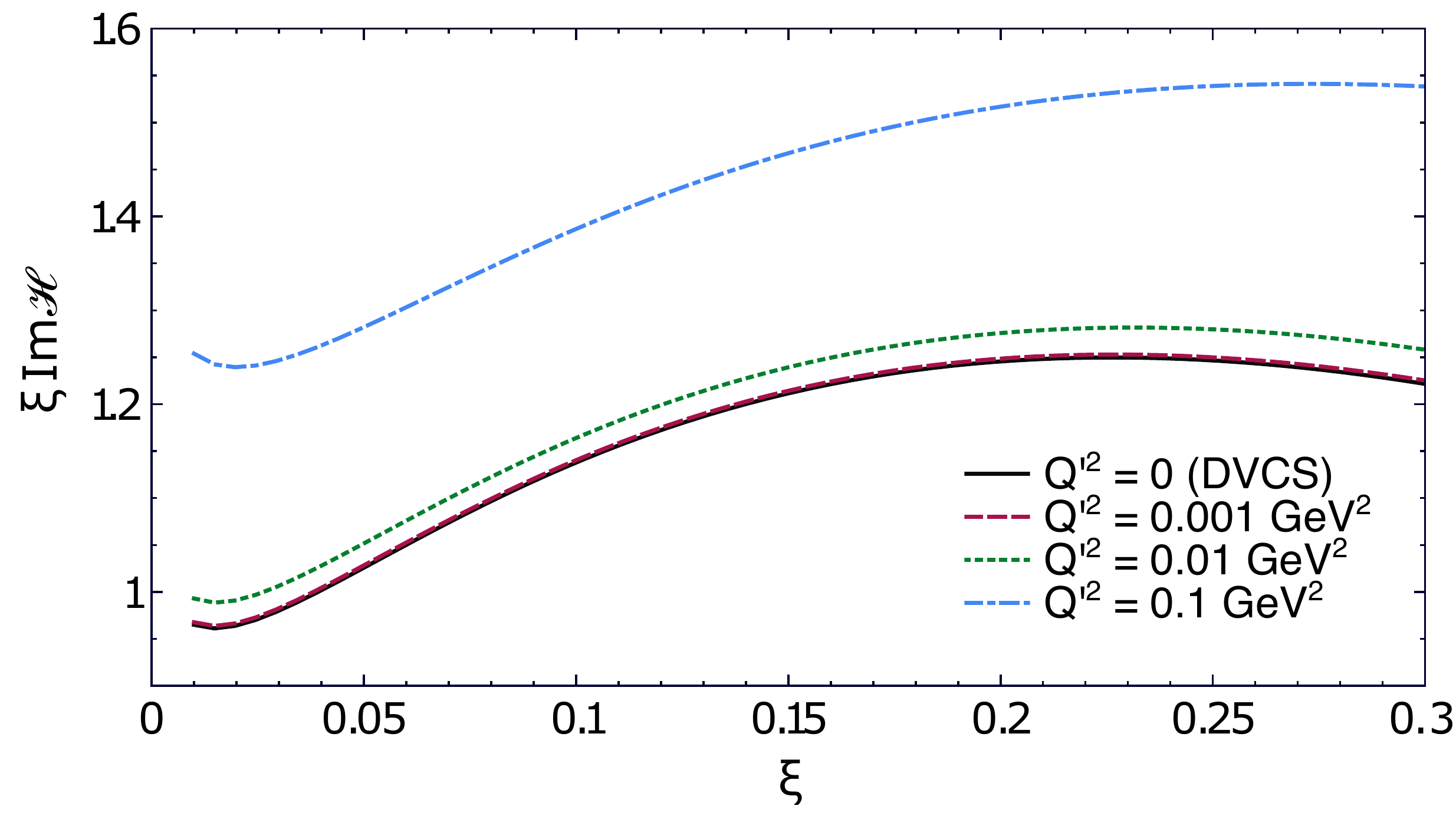}
    \includegraphics[width=0.45\textwidth]{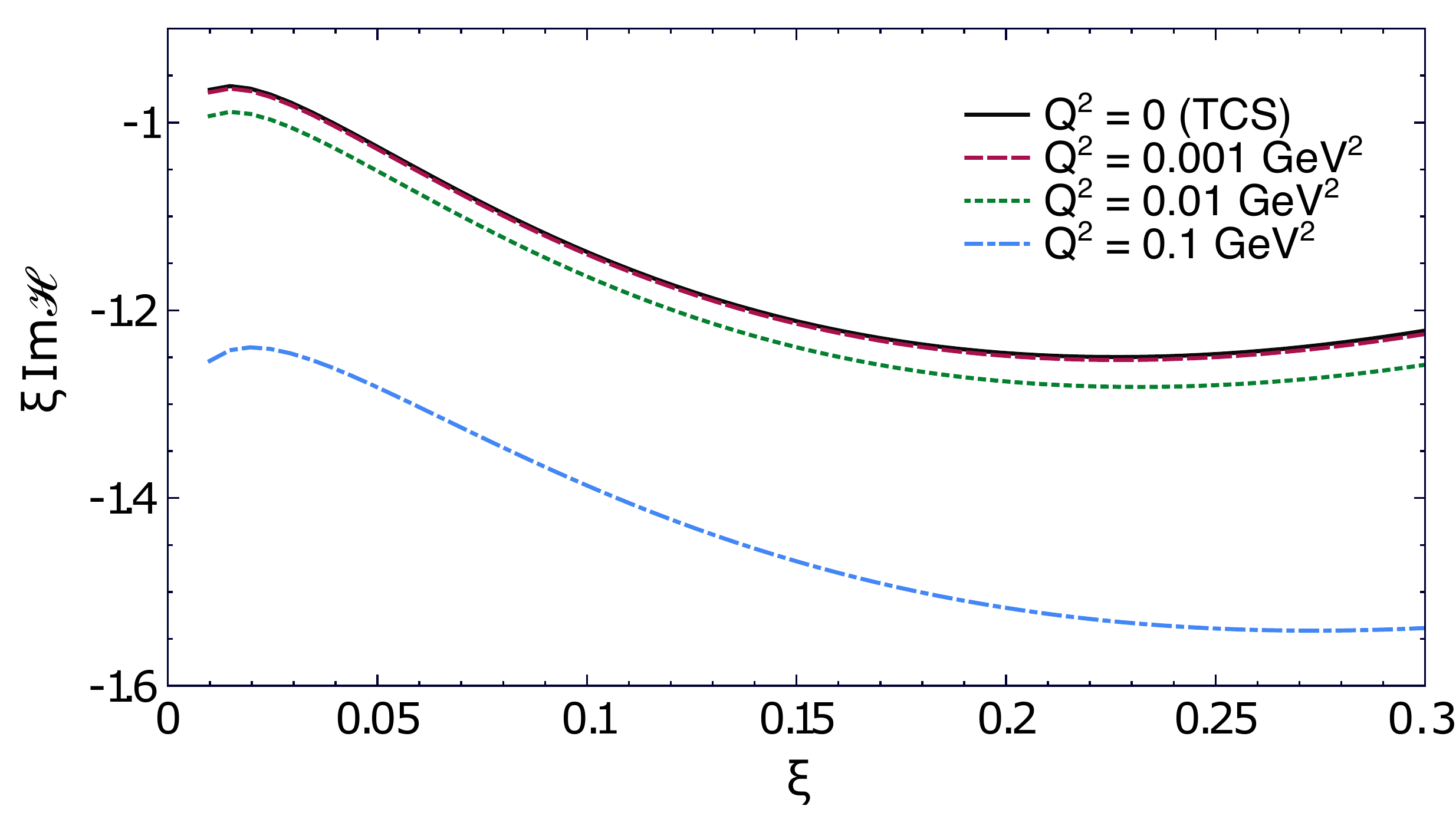}
    \caption{Imaginary part of DDVCS Compton form factor $\mathcal{H}$ as a function of $\xi$ approaching (left) DVCS and (right) TCS limits. The curves for DVCS and TCS limits, corresponding to $Q'^2 = 0$ and $Q^2 = 0$ values, respectively, are obtained with independent codes. Note that these curves nearly overlap with those for $Q'^2 = 0.001\ \mathrm{GeV}^2$ and $Q^2 = 0.001\ \mathrm{GeV}^2$. The left (right) plot is for $Q^2 = 1.5\ \mathrm{GeV}^2$ ($Q'^2 = 1.5\ \mathrm{GeV}^2$) and $t = -0.15\ \mathrm{GeV}^2$.}
    \label{fig:limitCFFAsXi}
\end{figure}

In Fig.~\ref{fig:limitCFFAsXi} we show how DDVCS CFFs plotted as a function of $\xi$ evolve as $Q^2 \to 0$ and $Q'^2 \to 0$. The curves for limits, $Q^2 = 0$ and $Q'^2 = 0$, are obtained with independent codes for DVCS and TCS processes available in PARTONS. One can conclude that the limits are reached without any discontinuities, hence DDVCS CFFs exhibit the proper reduction to DVCS and TCS counterparts when one of the two virtualities goes to zero. Although the presented quantity is only the imaginary part of CFF $\mathcal{H}$,  figures for real parts and other CFFs (not shown in this manuscript) lead to the same conclusions.
    
\begin{figure}[!ht]
    \centering
    \includegraphics[width=0.45\textwidth]{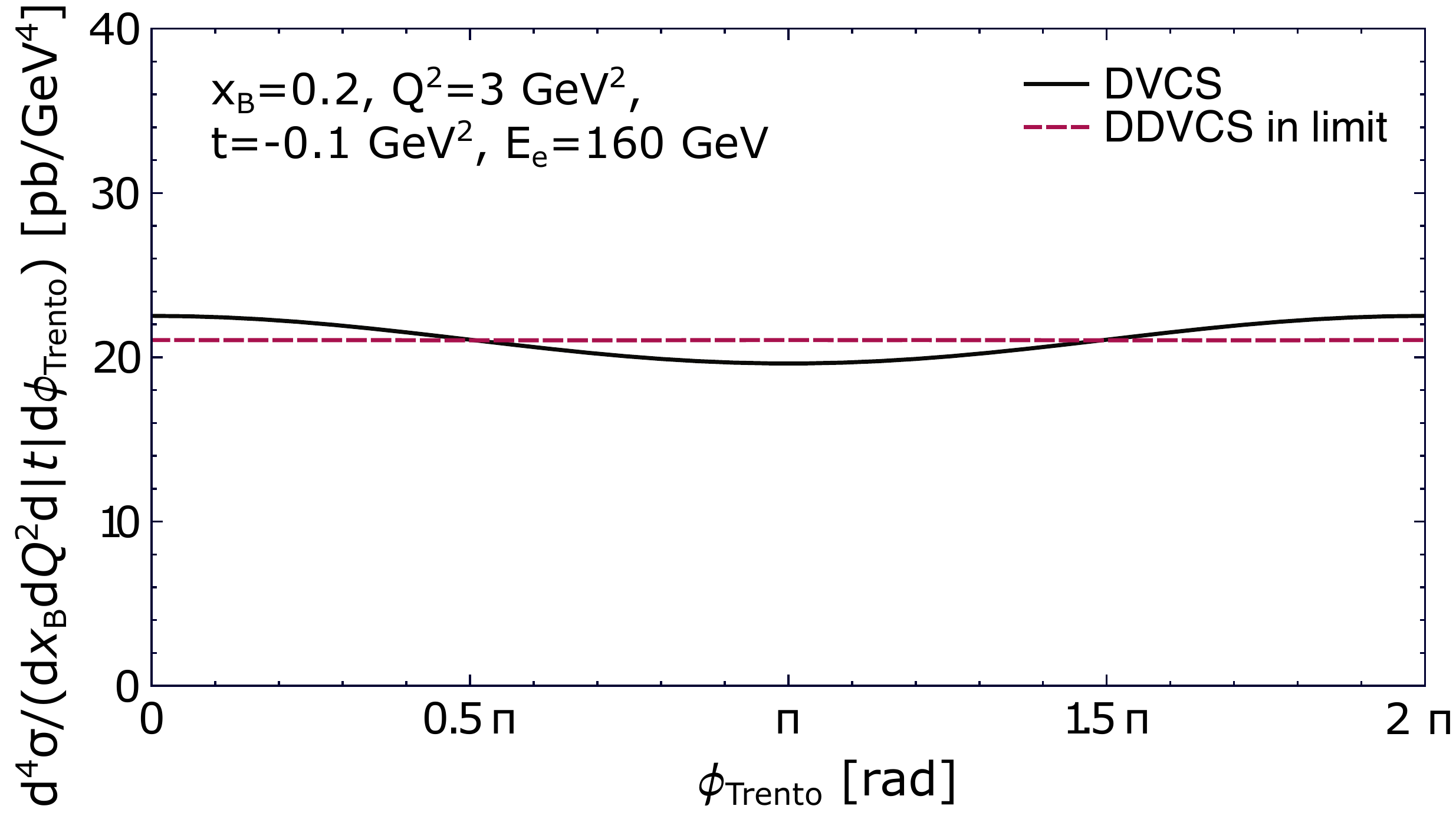}
    \includegraphics[width=0.45\textwidth]{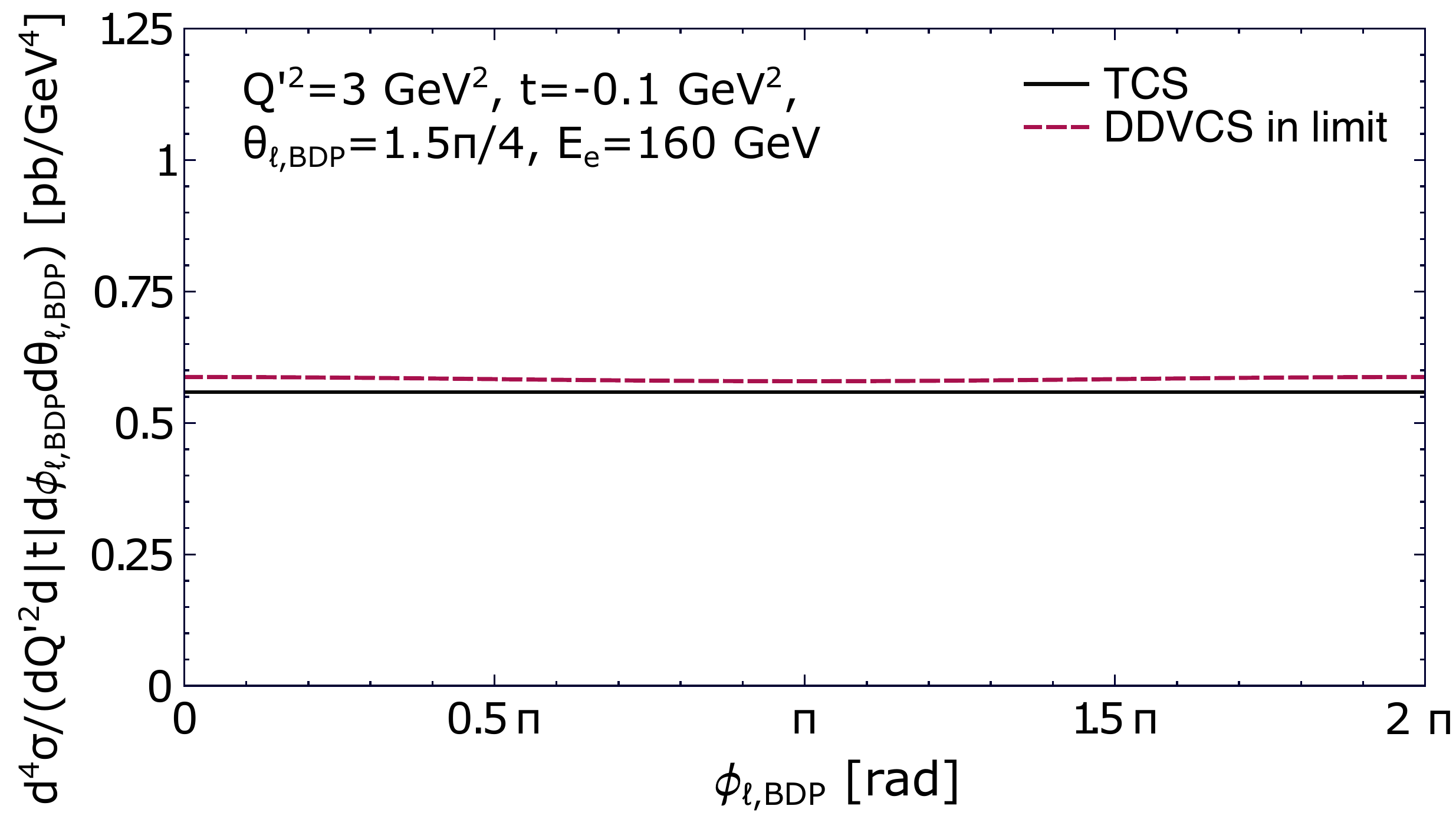}
    \includegraphics[width=0.45\textwidth]{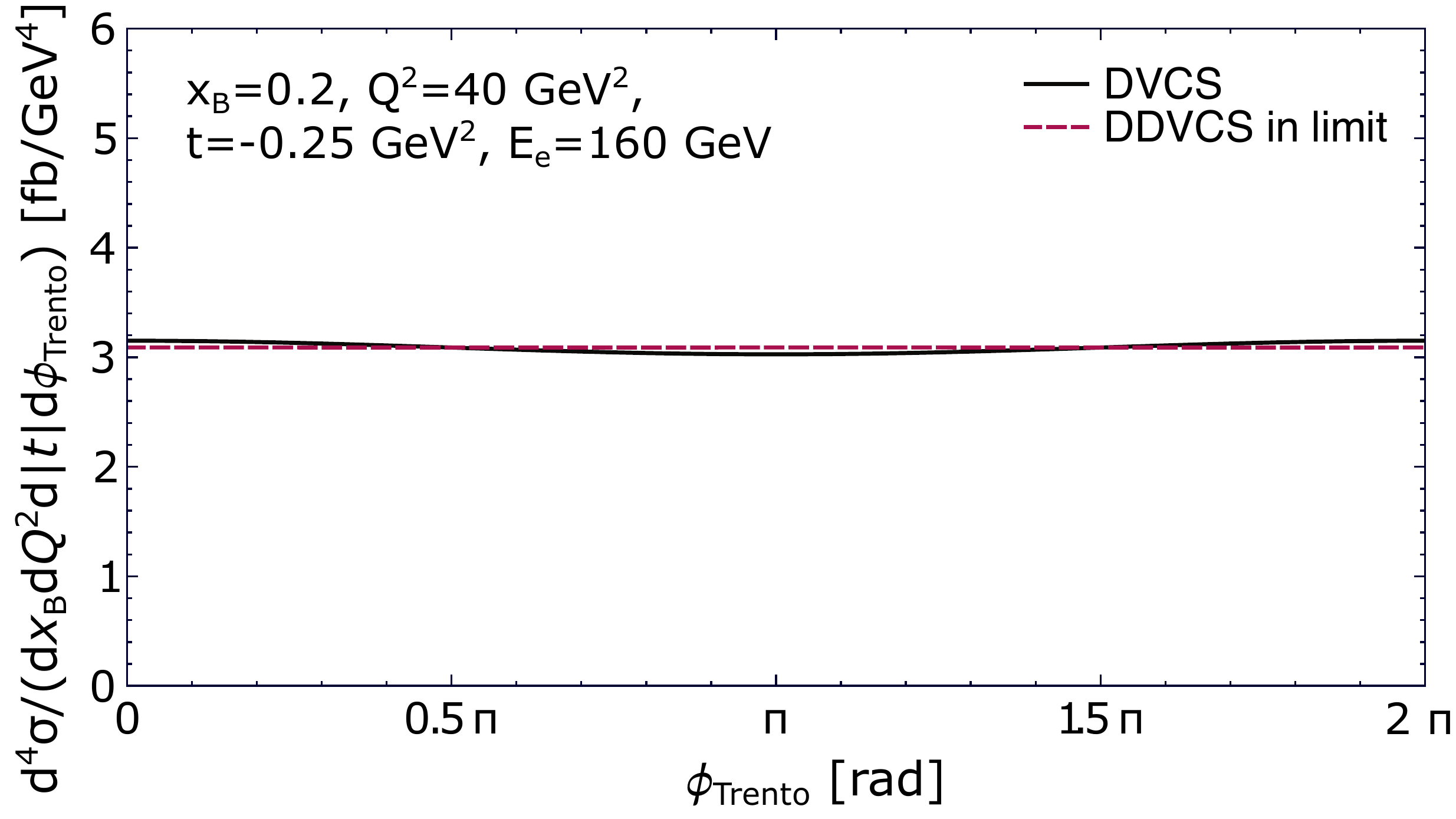}
    \includegraphics[width=0.45\textwidth]{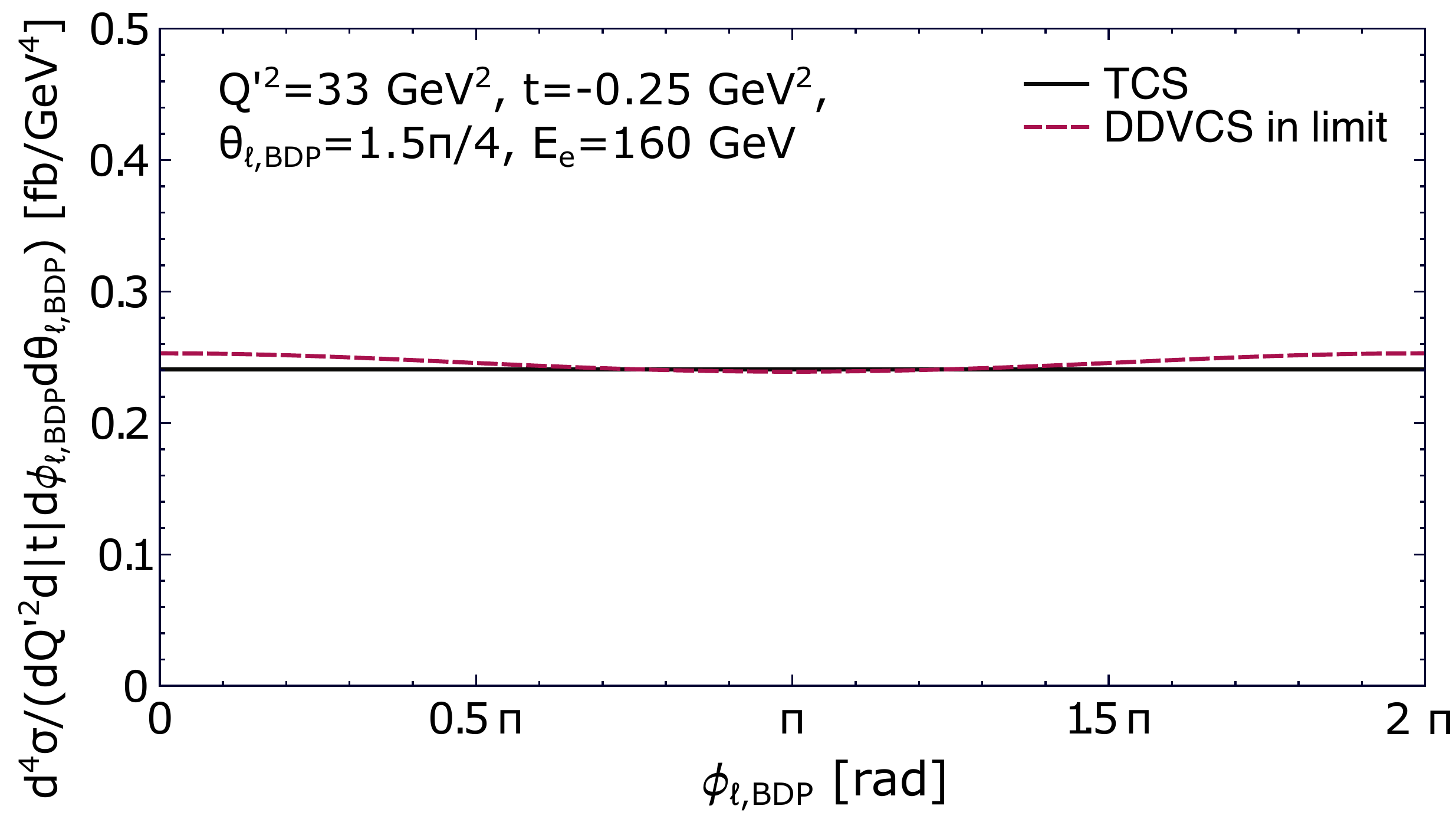}
    \caption{Comparison of DDVCS and (left) DVCS and (right) TCS cross-sections for pure VCS sub-process. Corresponding kinematic configurations are specified in the plots (all are for the fixed target). DDVCS cross-sections are modified according to Eqs. \eqref{eq:limitPrescriptionDVCS} and \eqref{eq:limitPrescriptionTCS}. Those for DVCS and TCS are evaluated with independent codes.}
    \label{fig:limitCSDDVCS}
\end{figure}

\begin{figure}[!ht]
    \centering
    \includegraphics[width=0.45\textwidth]{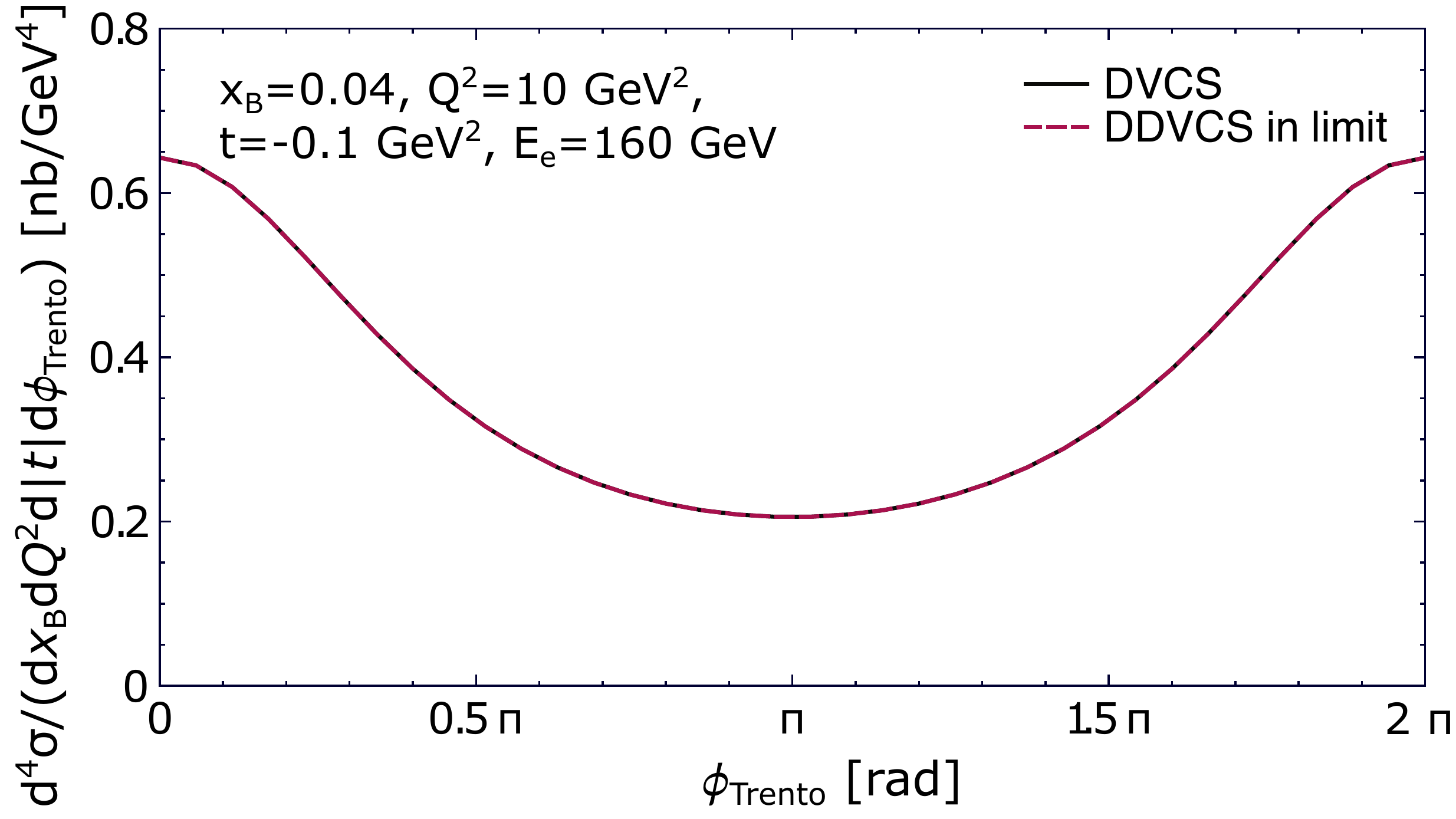}
    \includegraphics[width=0.45\textwidth]{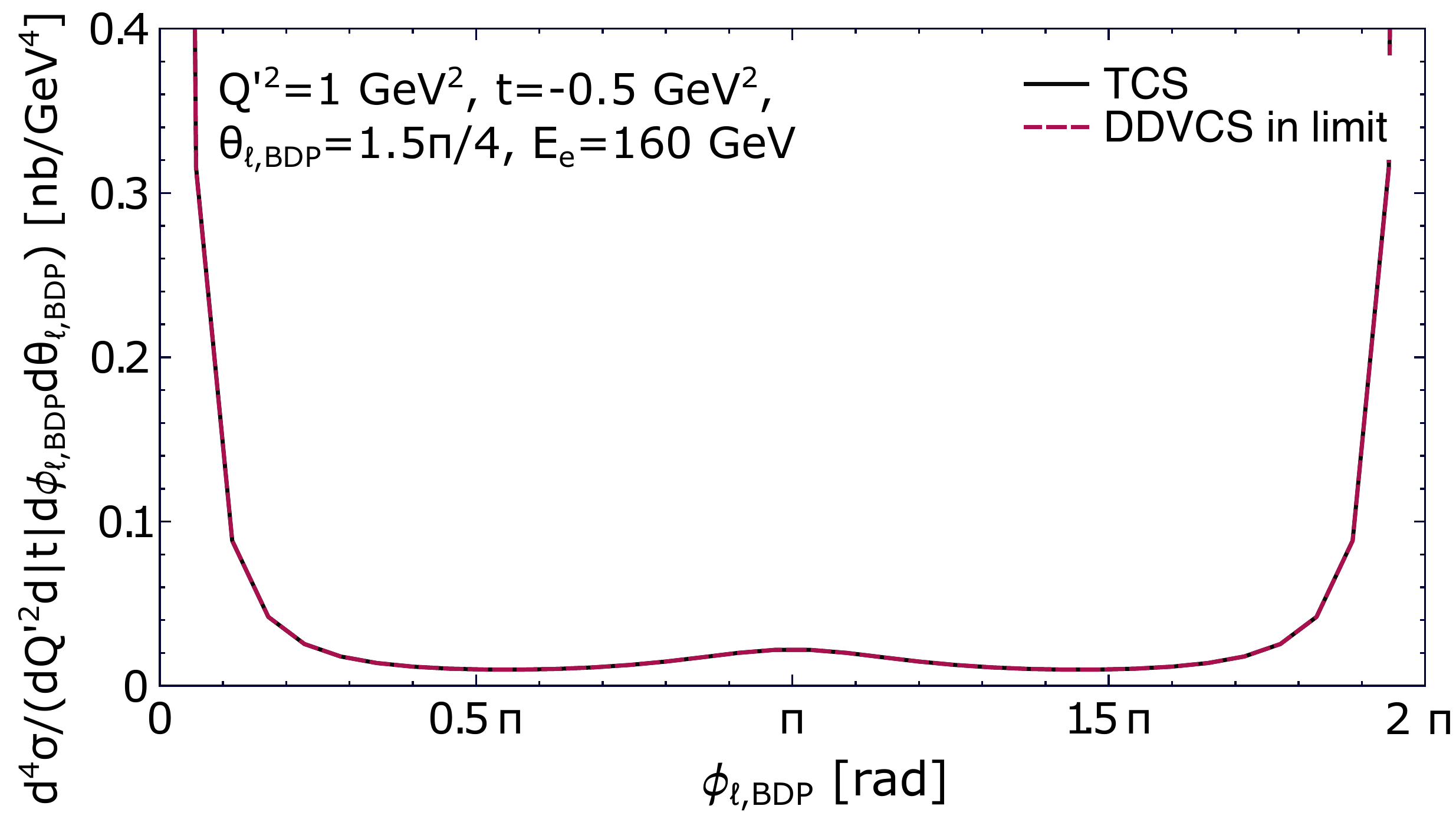}
    \caption{The same as Fig. \ref{fig:limitCSDDVCS} but for BH sub-process.}
    \label{fig:limitCSBH}
\end{figure}

The comparison for cross-sections is shown in Fig.~\ref{fig:limitCSDDVCS} for pure VCS sub-processes and in Fig.~\ref{fig:limitCSBH} for BH. Also here DVCS and TCS limits are evaluated with independent codes available in PARTONS. These codes are numerical implementations of works published in Refs. \cite{BELITSKY2014214} and \cite{Berger:2001xd}. 

For pure VCS sub-processes shown in Fig.~\ref{fig:limitCSDDVCS} the comparison with the limits is presented for two kinematic configurations, which only differ by either $|t|/Q^2$ or $|t|/Q'^2$ ratios. We see that the relative difference between pure DDVCS and the limits is reduced as these ratios become smaller. This signals that the observed differences stem from kinematic higher-twist corrections, which are related to the choice of the frame used to describe a given process. The effect is expected, as DVCS and TCS are described in fixed target frames where virtual photons move along the $z$-axis. With two virtual photons in DDVCS case the frame must be different, resulting in differences in the twist expansion, see Sect.~\ref{sectAmp} for more details. As for BH we do not deal with this type of expansion, the agreement with DVCS and TCS limits is exact, as demonstrated in Fig. \ref{fig:limitCSBH}. 

\section{DDVCS observables}
\label{sectObs}

\begin{figure}[!ht]
    \centering
    \includegraphics[width=0.31\textwidth]{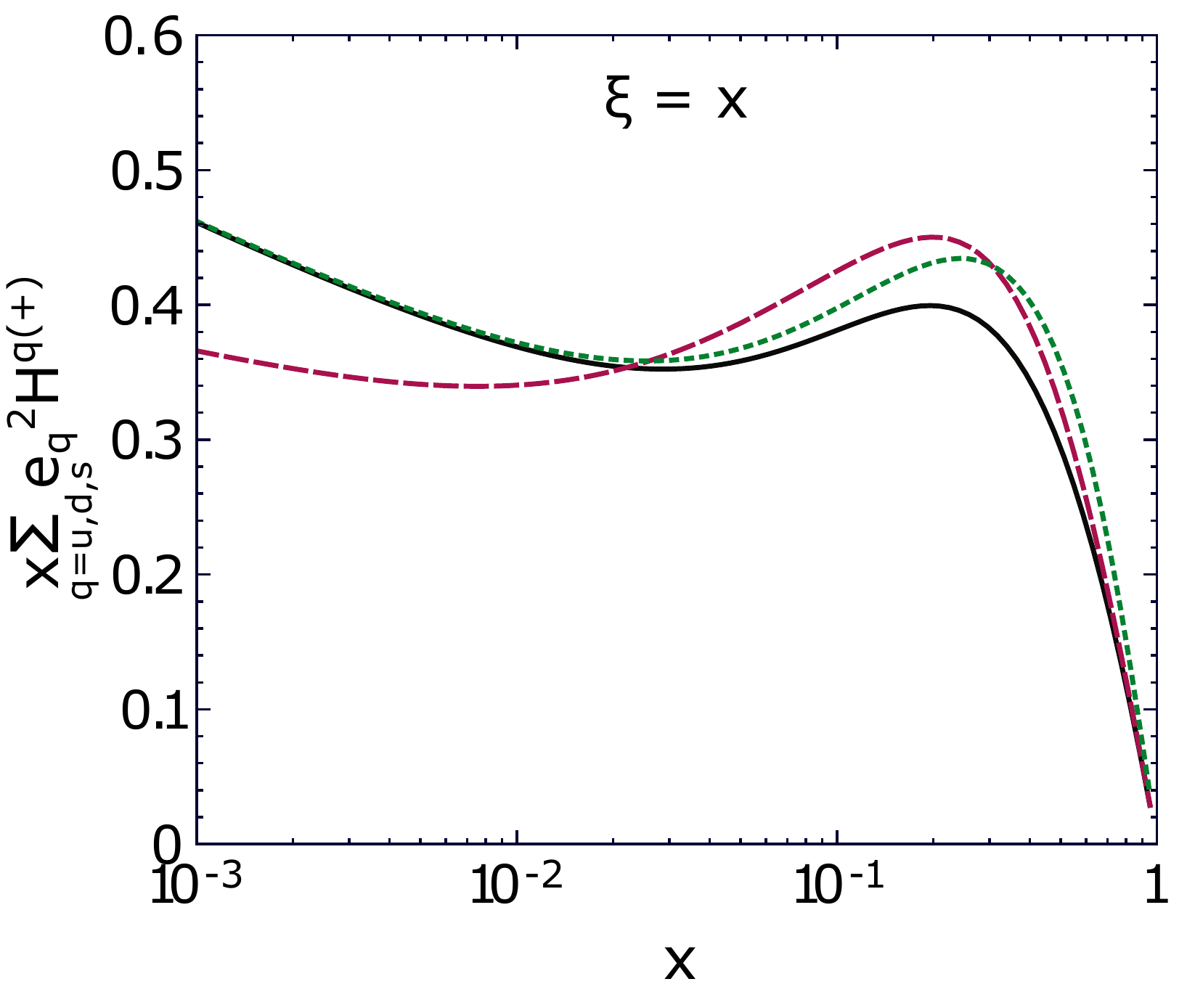}
    \includegraphics[width=0.31\textwidth]{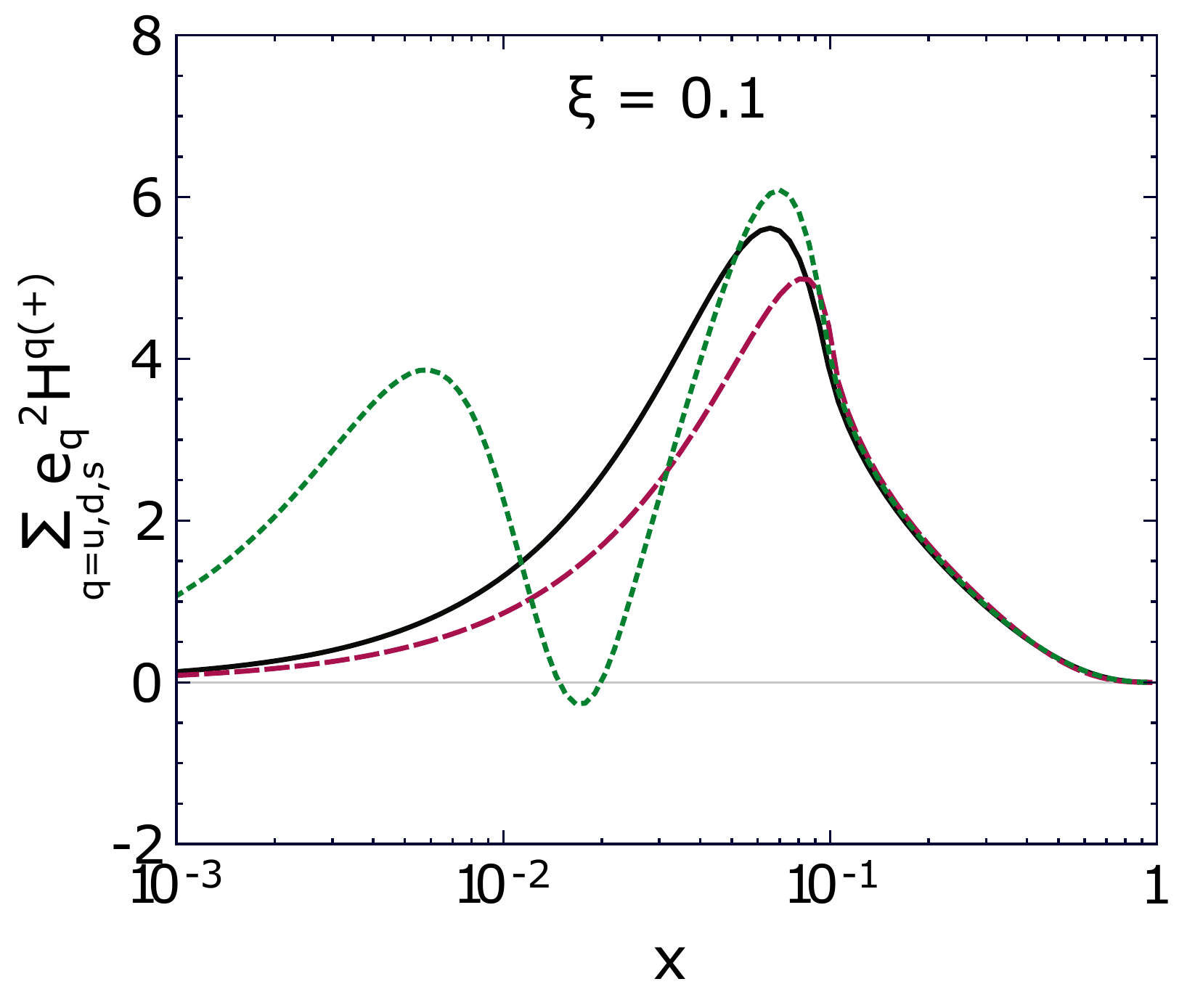}
    \includegraphics[width=0.31\textwidth]{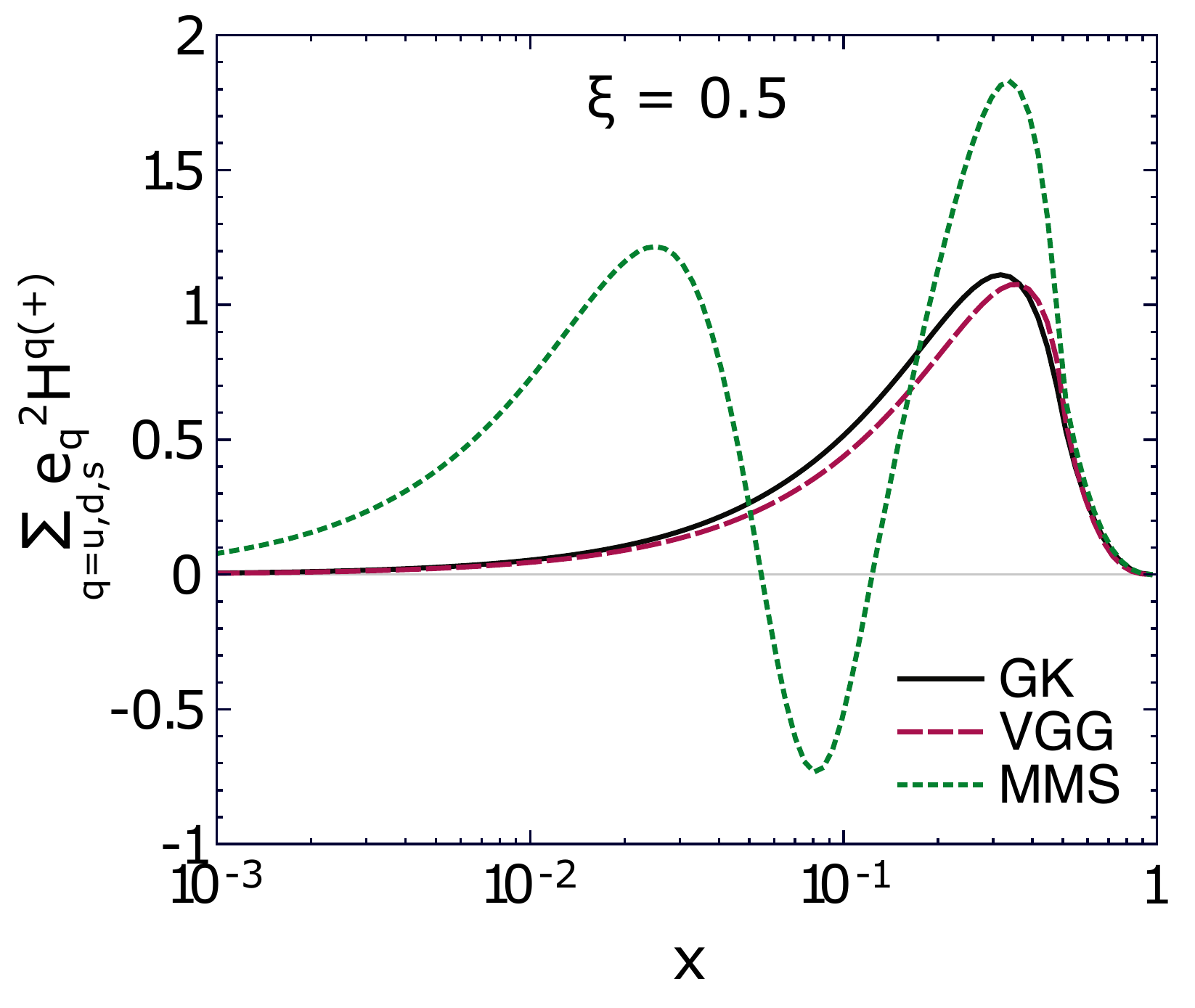}
    \caption{Distributions of $\sum_q e_{q}^{2}H^{q(+)}(x, \xi, t)$ at $t=-0.1~\mathrm{GeV}^{2}$, where $q=u,d,s$ flavours for (left) $\xi = x$, (middle) $\xi = 0.1$ and (right) $\xi = 0.5$. The solid black, dashed red and dotted green curves describe the GK, VGG and MMS GPD models, respectively. The C-even part of a given vector GPD is defined as: $H^{q(+)}(x, \xi, t) = H^{q}(x, \xi, t) - H^{q}(-x, \xi, t)$. The  scale is choosen as $\mu_{F}^{2} = 4~\mathrm{GeV}^{2}$.}
    \label{fig:gpds}
\end{figure}

In this section we present selected DDVCS observables in the kinematics of current and future experiments. We stress the usefulness of this process to constrain various types of GPDs by checking how DDVCS can be used to distinguish between different GPD models giving otherwise similar predictions for DVCS and TCS. For these purposes we use the GK \cite{Goloskokov_2007, Goloskokov_2007_2}, VGG \cite{guichon1998, guichon1999, Goeke_2001, guidal2005} and MMS \cite{Mezrag:2013mya} GPD models implemented in PARTONS. We note that the MMS model only differs from the GK one by the valence part described by either one-component \cite{Radyushkin:2011dh} (MMS) or two-component \cite{Radyushkin:1998es, Radyushkin:1998bz} (GK) double distributions. This difference in the choice of the double distribution makes MMS unique also with respect to VGG, and is responsible for a different behaviour of the model in $x \neq \xi$ domain. The three models are depicted in Fig.~\ref{fig:gpds} for the dominant distribution probed by DDVCS at leading order, $\sum_{q={\{u,d,s\}}} e_{q}^{2}H^{q(+)}(x, \xi, t)$, where $H^{q(+)}(x, \xi, t) = H^{q}(x, \xi, t) - H^{q}(-x, \xi, t)$. We note that all three models are similar in the DGLAP region ($|x| > \xi$), which is a consequence of the common PDF limit and a similar modelling method, but they differ significantly in the ERBL region ($|x| < \xi$). The latter one is directly probed by DDVCS at LO, making this process a convenient tool to distinguish between such various GPD models. 

The selected observables are unpolarized differential cross-sections:
\begin{equation}
    \sigma_{UU}(\phiLBDP) = \int_0^{2\pi} d\phi\int_{\pi/4}^{3\pi/4}d\thetaLBDP\ \sin\thetaLBDP \left( \frac{d^7\sigma^{\rightarrow}}{dx_B dQ^2 dQ'^2 d|t| d\phi d\OmegaLBDP} + \frac{d^7\sigma^{\leftarrow}}{dx_B dQ^2 dQ'^2 d|t| d\phi d\OmegaLBDP} \right)\,,
    \label{eq:ass1}
\end{equation}
and their cosine components:
\begin{equation}
\sigma_{UU}^{\cos(n \phiLBDP)}(\phiLBDP) = M_{UU}^{\cos(n \phiLBDP)}\cos(n \phiLBDP) \,.
\end{equation}
where
\begin{equation}
M_{UU}^{\cos(n \phiLBDP)} = \frac{1}{N}\int_0^{2\pi} d\phiLBDP \cos(n \phiLBDP)  \sigma_{UU}(\phiLBDP)  \,.
\end{equation}
Here, $N=2\pi$ for $n = 0$ and $N=\pi$ for $n > 0$.

We also consider asymmetries for longitudinally polarized electron beam: 
\begin{equation}
    A_{LU}(\phiLBDP) = \frac{\Delta\sigma_{LU}(\phiLBDP)}{\sigma_{UU}(\phiLBDP)}\,,
    \label{eq:ass2}
\end{equation}
where
\begin{equation}
    \Delta\sigma_{LU}(\phiLBDP) = \int_0^{2\pi} d\phi\int_{\pi/4}^{3\pi/4}d\thetaLBDP\ \sin\thetaLBDP \left( \frac{d^7\sigma^{\rightarrow}}{dx_B dQ^2 dQ'^2 d|t| d\phi d\OmegaLBDP} - \frac{d^7\sigma^{\leftarrow}}{dx_B dQ^2 dQ'^2 d|t| d\phi d\OmegaLBDP} \right)\,.
    \label{eq:ass3}
\end{equation}
Here, we omit the dependence on variables other than angles, while right and left arrows stand for positive and negative helicity of the incoming electron beam, respectively. In order to reduce the contribution coming from the pure BH sub-process, the integration over $\thetaLBDP$ angle is performed in the limited range $(\pi/4, 3\pi/4)$. The cross-section difference $\Delta\sigma_{LU}(\phiLBDP)$ is sensitive to $\sin \phiLBDP$ part of the interference, and therefore carries information on the imaginary part of CFFs \cite{Belitsky:2003fj}. If one neglects the $\phiLBDP$-dependence of $\sigma_{UU}(\phiLBDP)$ the same interpretation applies to the asymmetry $A_{LU}(\phiLBDP)$. This asymmetry in the $Q^2 \to 0$ limit can be related to the circular asymmetry in TCS~\cite{CLAS:2021lky, Grocholski_2020}. In the following we do not show results for $A_{LU}(\phi)$, i.e. the asymmetry obtained with $\phiLBDP \leftrightarrow \phi$ replacement in Eqs.~\eqref{eq:ass1}-\eqref{eq:ass3}, as its magnitude in the considered kinematics is much smaller than that for $A_{LU}(\phiLBDP)$ being a consequence of $Q'^2 \gg Q^2$. 

\begin{figure}[!ht]
    \centering
    \includegraphics[width=0.45\textwidth]{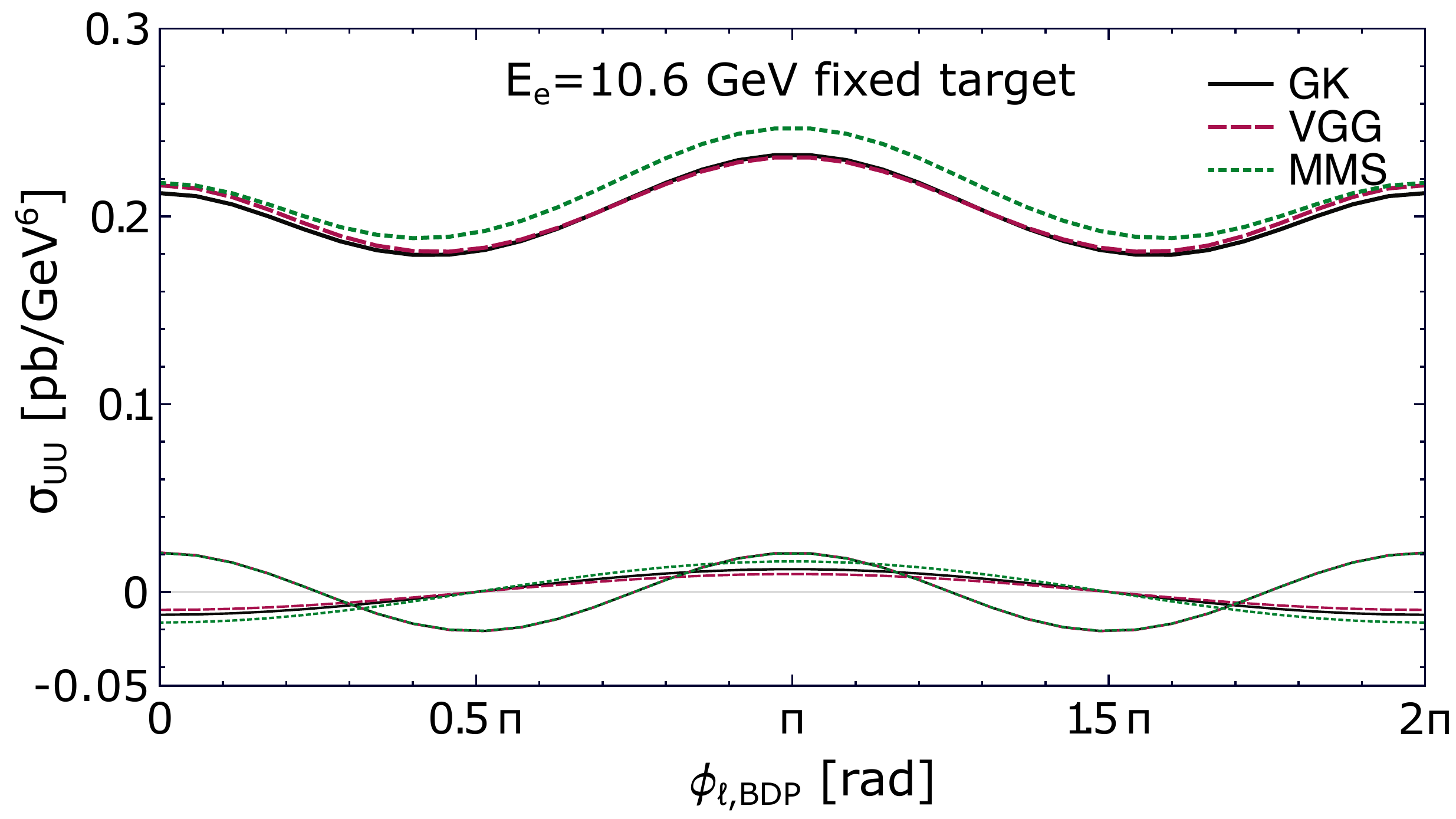}
    \includegraphics[width=0.45\textwidth]{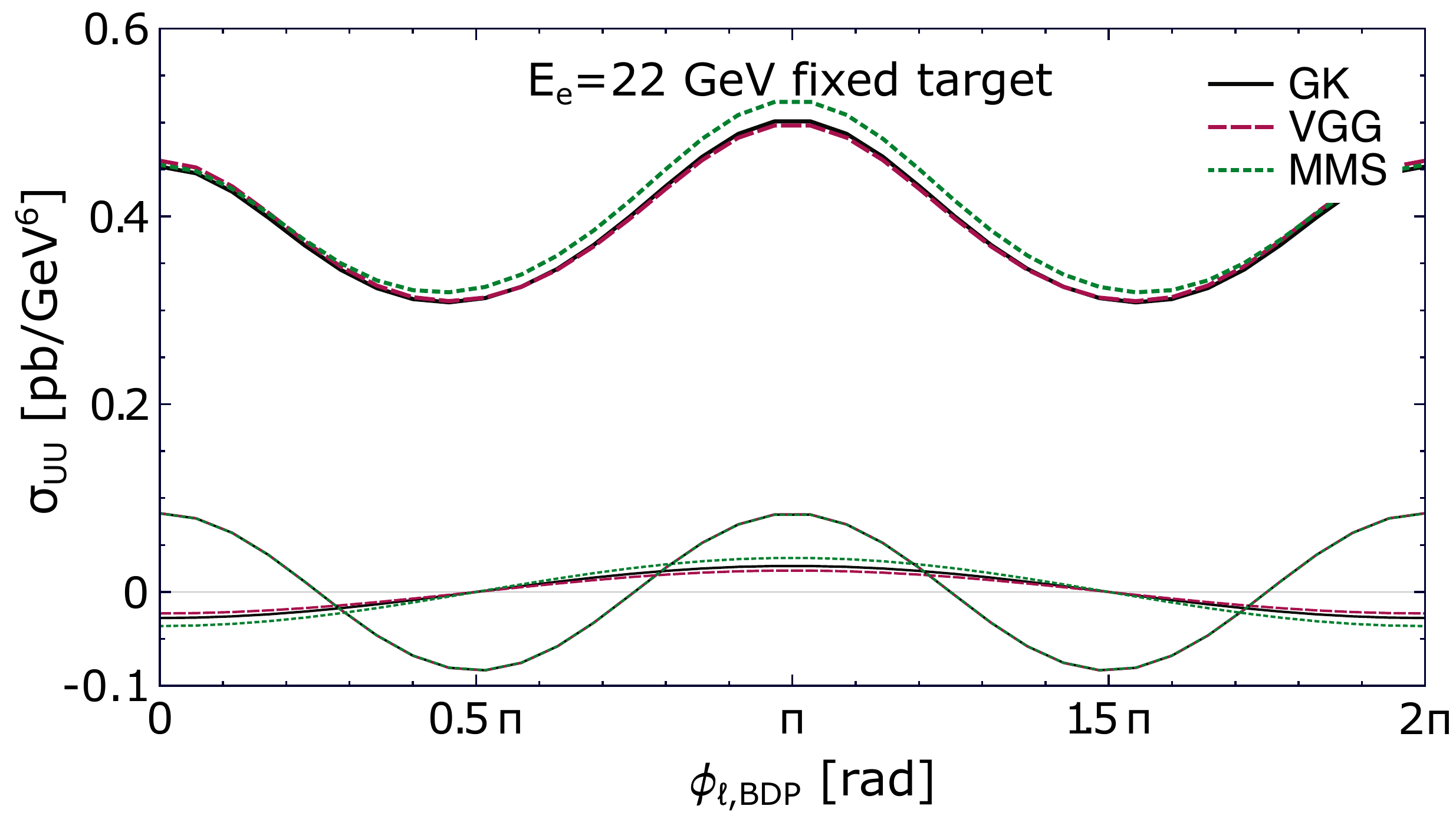}
    \includegraphics[width=0.45\textwidth]{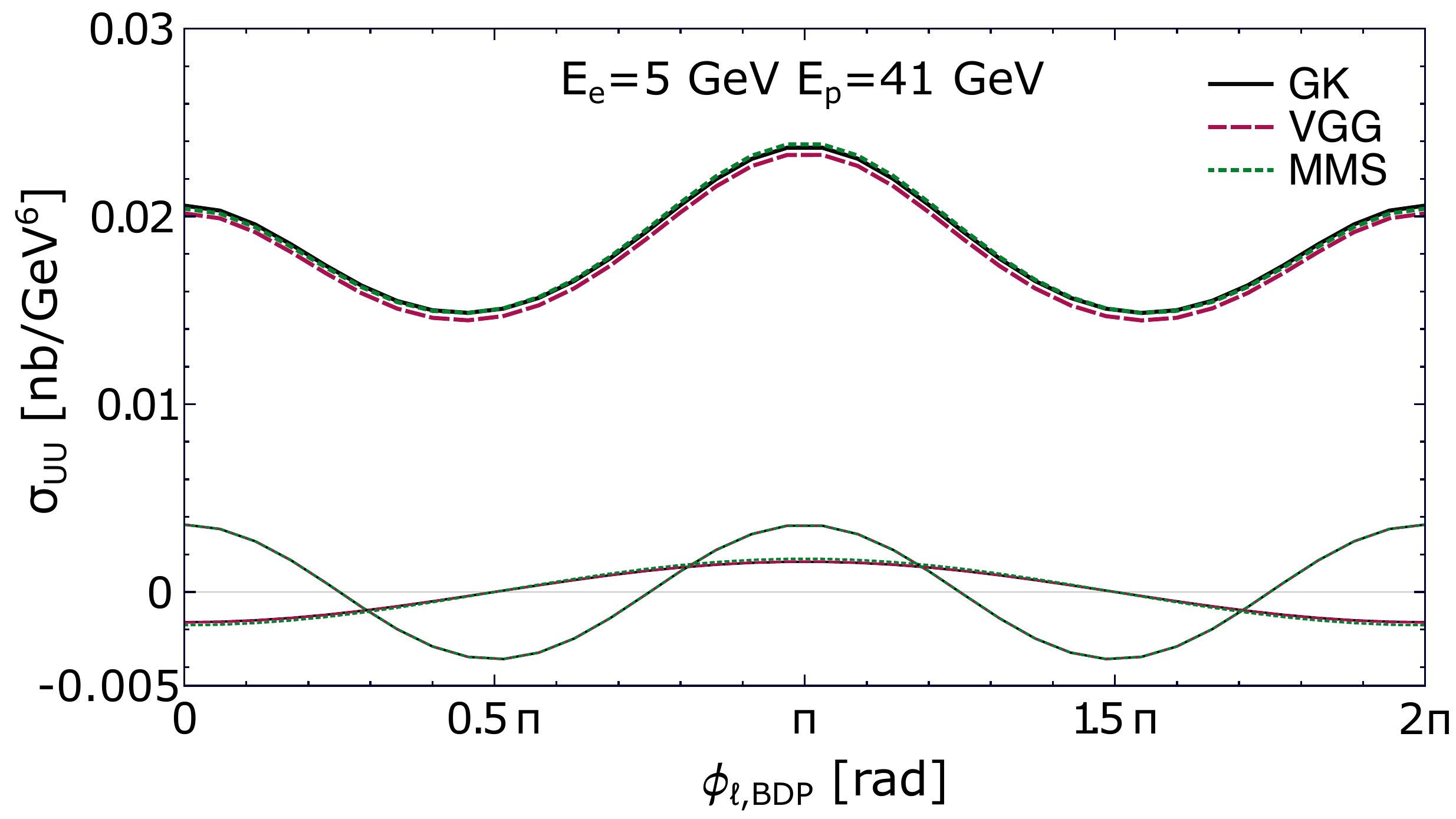}
    \includegraphics[width=0.45\textwidth]{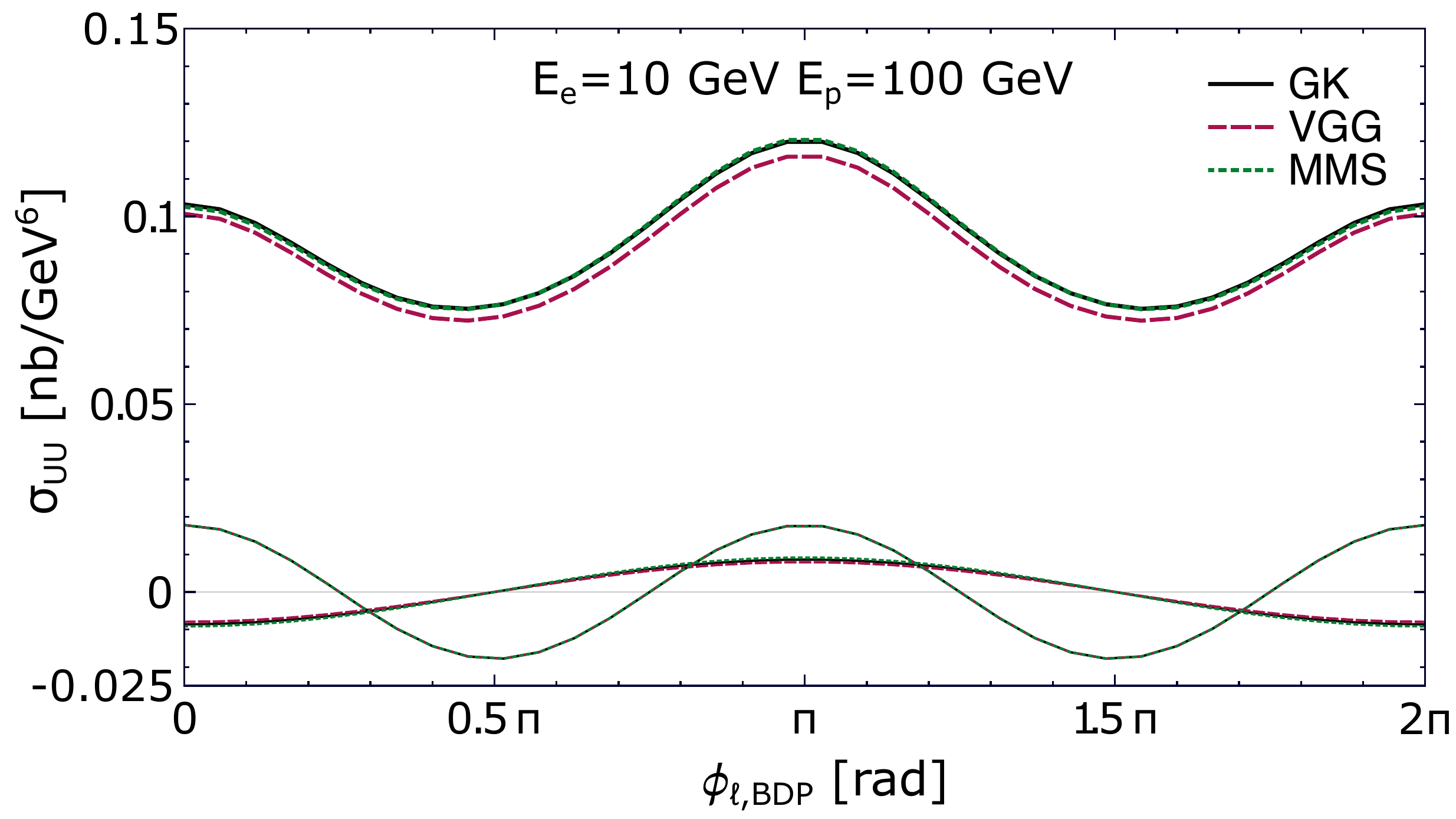}
    \caption{Unpolarized cross-section, $\sigma_{UU}(\phiLBDP)$, and its  $\sigma_{UU}^{\cos\phiLBDP}(\phiLBDP)$ and $\sigma_{UU}^{\cos2\phiLBDP}(\phiLBDP)$ components for beam energies specified in the plots and extra kinematic conditions given in Table~\ref{tab:typicalValuesJLabEIC}. The solid black, dashed red and dotted green curves are for GK, VGG and MMS GPD models, respectively.}
    \label{fig:denominators}
\end{figure}

\begin{figure}[!ht]
    \centering
    \includegraphics[width=0.45\textwidth]{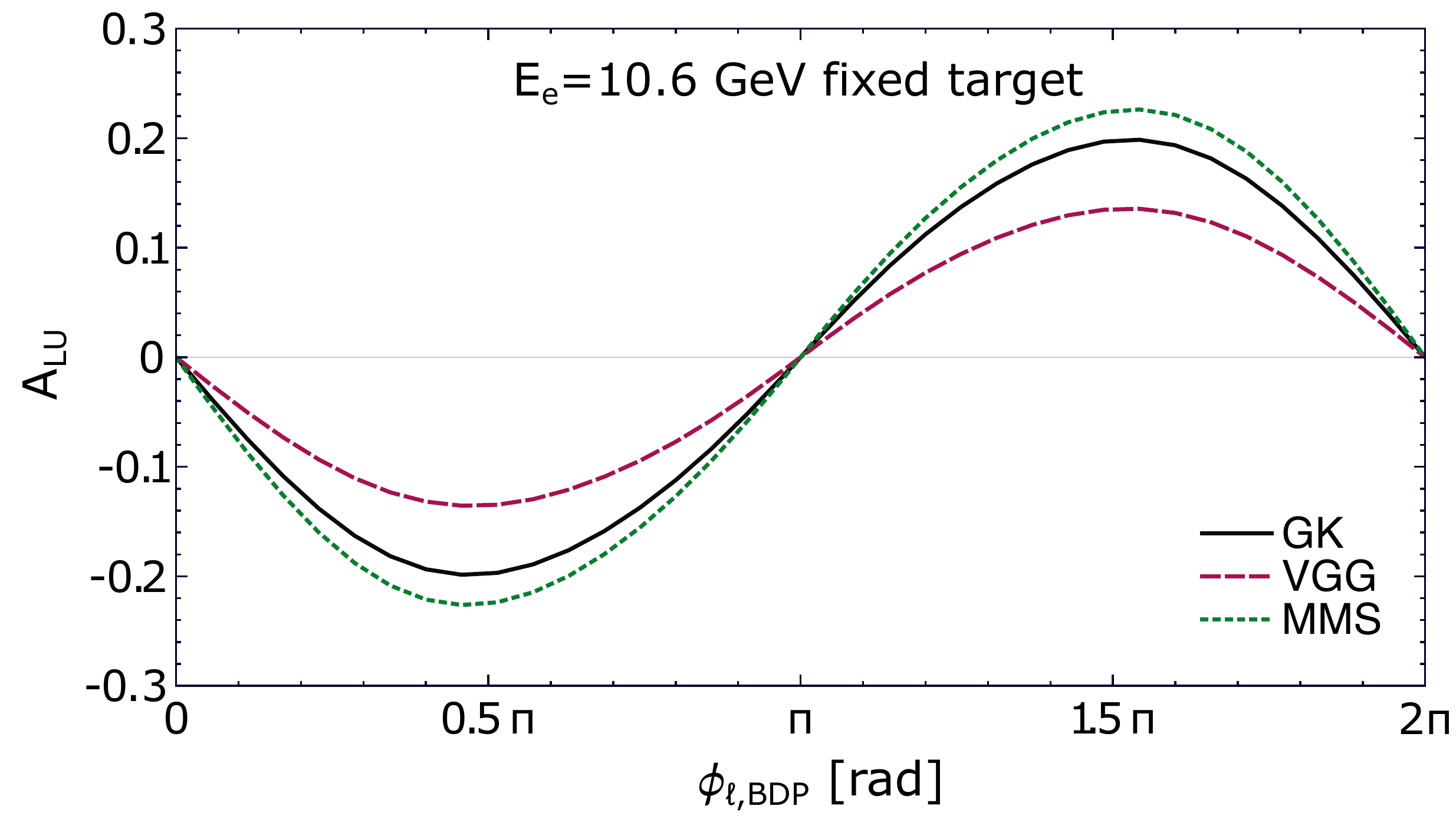}
    \includegraphics[width=0.45\textwidth]{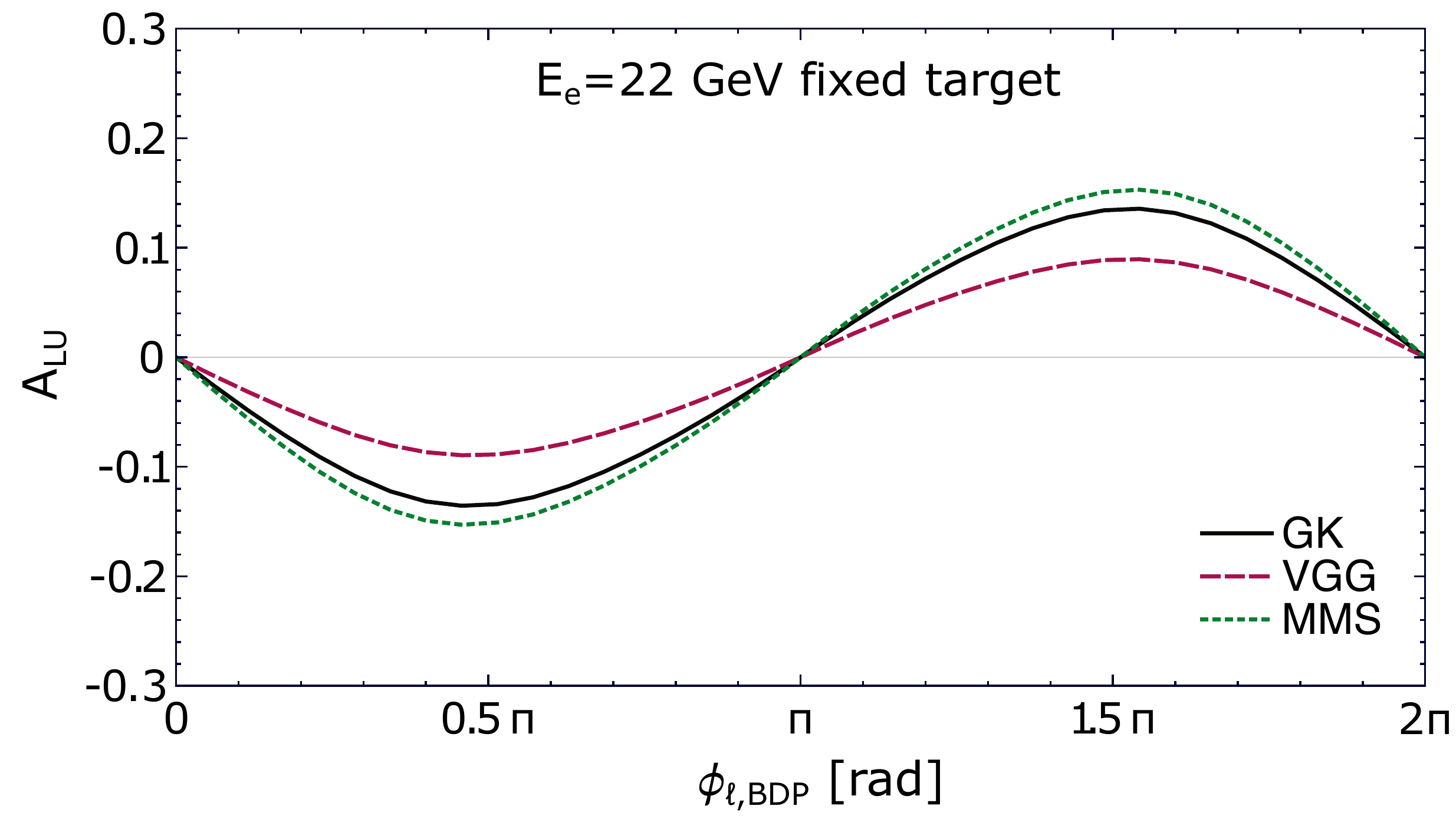}
    \includegraphics[width=0.45\textwidth]{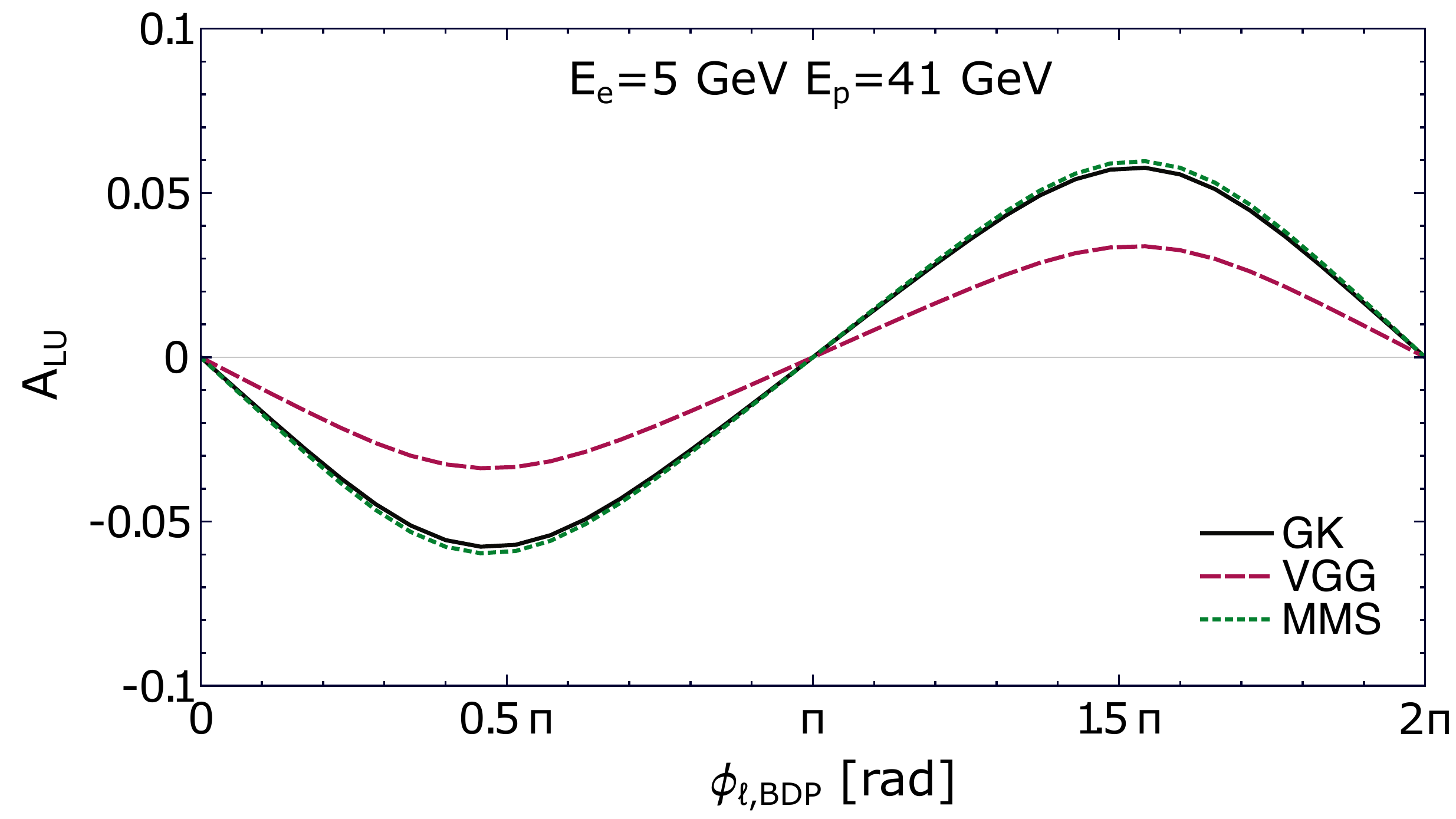}
    \includegraphics[width=0.45\textwidth]{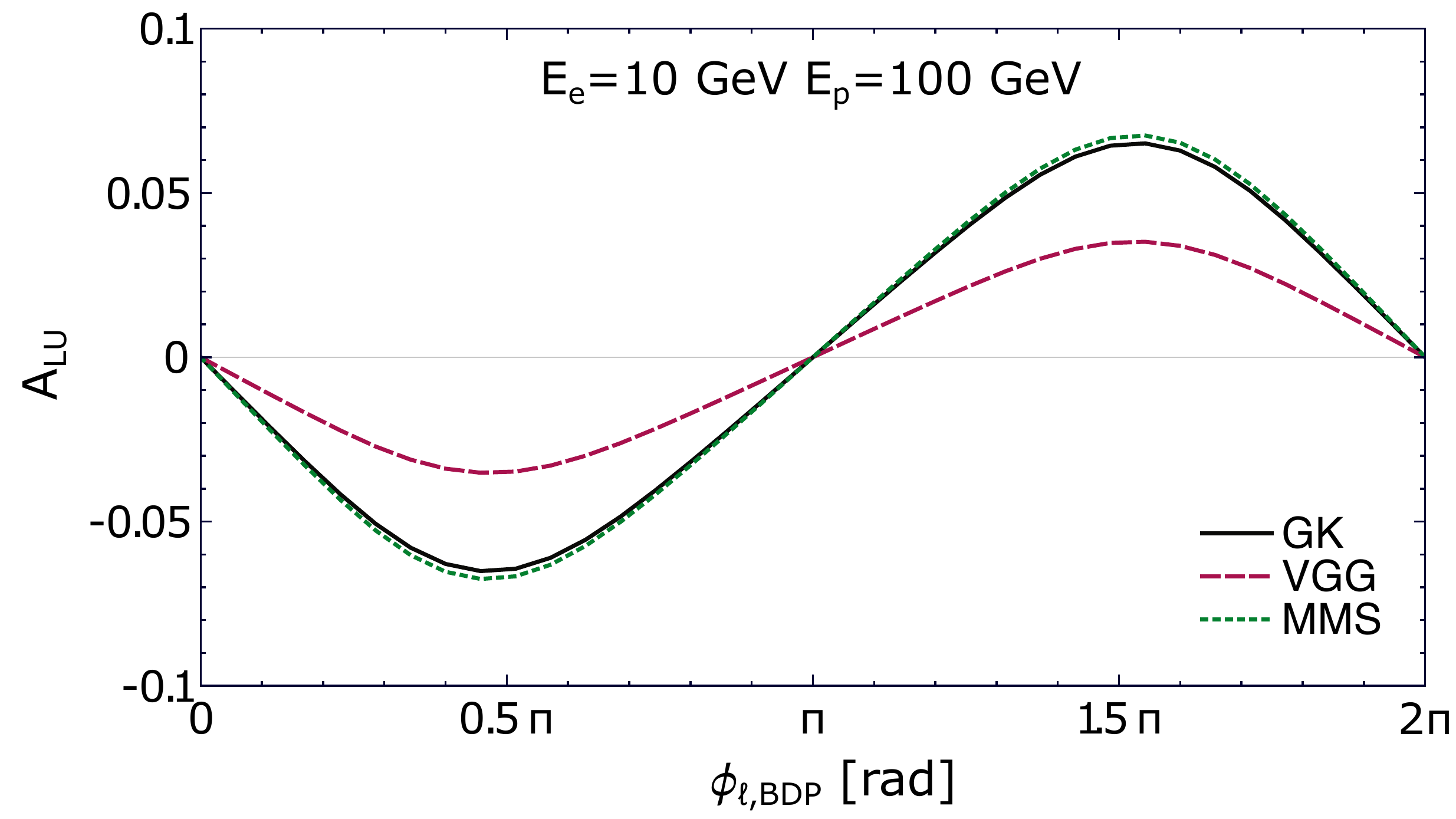}
    \caption{Asymmetry $A_{LU}(\phiLBDP)$ for beam energies specified in the plots and extra kinematic conditions given in Table~\ref{tab:typicalValuesJLabEIC}. The solid black, dashed red and dotted green curves are for GK, VGG and MMS GPD models, respectively.}
    \label{fig:asymetries}
\end{figure}

The predictions for JLab12, JLab20+ and EIC experiments are shown in Fig.~\ref{fig:denominators} for unpolarized cross-sections and their cosine components, and in Fig.~\ref{fig:asymetries} for the asymmetries. The plotted quantities are evaluated at kinematics specified in Table~\ref{tab:typicalValuesJLabEIC}. In our study the timelike virtuality, $Q'^2$, has been taken large enough to avoid resonances, as suggested by the recent measurement of TCS by CLAS \cite{CLAS:2021lky}. The selection of spacelike virtuality, $Q^2$, allows one to probe $|\rho|$ significantly smaller than $\xi$, still keeping a reasonable cross-section of DDVCS. Values of the inelasticity variable, $y$, have been taken in accordance to Monte Carlo simulations presented in Sect.~\ref{sectMC} and allow to achieve good statistics of collected events in the corresponding experiment. 

\begin{table}[!ht]
\begin{ruledtabular}
\begin{tabular}{@{}lccccc@{}}
Experiment & Beam energies & $y$ & $|t|$ & $Q^2$ & $Q'^2$ \\ 
& $[\mathrm{GeV}]$ &  & $[\mathrm{GeV^2}]$ & $[\mathrm{GeV^2}]$ & $[\mathrm{GeV^2}]$ \\[5pt] 
JLab12 & $E_{e} = 10.6$, $E_p = M$ & $0.5$ & $0.2$ & $0.6$ & $2.5$\\
JLab20+ & $E_{e} = 22$, $E_p = M$ & $0.3$ & $0.2$ & $0.6$ & $2.5$\\ 
EIC & $E_{e} = 5$, $E_p = 41$ & $0.15$ & $0.1$ & $0.6$ & $2.5$ \\
EIC & $E_{e} = 10$, $E_p = 100$ & $0.15$ & $0.1$ & $0.6$ & $2.5$\\
\end{tabular}
\end{ruledtabular}
\caption{DDVCS kinematics used for predictions of asymmetries presented in Fig.~\ref{fig:asymetries}.}
\label{tab:typicalValuesJLabEIC}
\end{table}

Figure~\ref{fig:denominators} for the unpolarized cross-sections show sizeable contributions of $\cos(n \phiLBDP)$ components for $n=0,1,2$. For the interpretation of these contributions we may use Ref.~\cite{Belitsky:2003fj}. The constant term, $\sigma_{UU}^{1}$, is dominated by BH, with a few percent contribution of pure DDVCS, mostly sensitive to the moduli of CFFs. The term $\sigma_{UU}^{\cos \phiLBDP}$ is particularly interesting, as it is induced by the interference between BH and pure DDVCS, and it carries information about the real parts of CFFs. Finally, $\sigma_{UU}^{\cos2 \phiLBDP}$ is only sensitive to the BH process. This term vanishes for $|\xi| = |\rho|$, i.e. is not observed in both DVCS and TCS limits.

Using Fig.~\ref{fig:asymetries} we conclude that the magnitude of the asymmetry $A_{LU}(\phiLBDP)$ is up to the order $20\%$ for JLab12, $15\%$ for JLab20+ and $3\%$-$7\%$ for EIC. Such sizable asymmetries and fairly large integrated cross-sections presented in Sect.~\ref{sectMC} strongly indicates the feasibility of meaningful DDVCS programmes at all considered facilities.


The measurability of the $R$ ratio \cite{Berger:2001xd} and the forward-backward asymmetry \cite{CLAS:2021lky} should also be carefully studied. These two observables have been specifically designed in former TCS studies  to take advantage of the charge asymmetry of the final muons to isolate the interference of the Bethe-Heitler and the TCS processes. In the DDVCS reaction, they probe the interference of the BH2 amplitude with the sum of the BH1 and DDVCS amplitudes.
\section{Monte Carlo studies}
\label{sectMC}

To make our study easy to use for the theory community, the cross-section formulae we obtained have been implemented in the open-source PARTONS framework \cite{Berthou:2015oaw}. The implementation of DDVCS in the EpIC Monte Carlo (MC) generator \cite{Aschenauer:2022aeb} has followed, making our work also accessible to experimentalists. This development is important to support the physics case of a new generation of experiments, like JLab12, JLab20+ and EIC, for which it is desirable to estimate the measurability of DDVCS. 

In this section we present first results obtained with EpIC for the DDVCS reaction, also checking the accuracy of this generator in reproducing the underlying cross-sections. Our results do not include any simulation of detector effects, and they are not affected by any efficiency one should take into account of in this kind of analysis. Therefore, the presented material should only be considered as a rough estimate and motivation for studying the measurability of DDVCS in more depth. 

\begin{figure}
    \centering
    \includegraphics[width=0.45\textwidth]{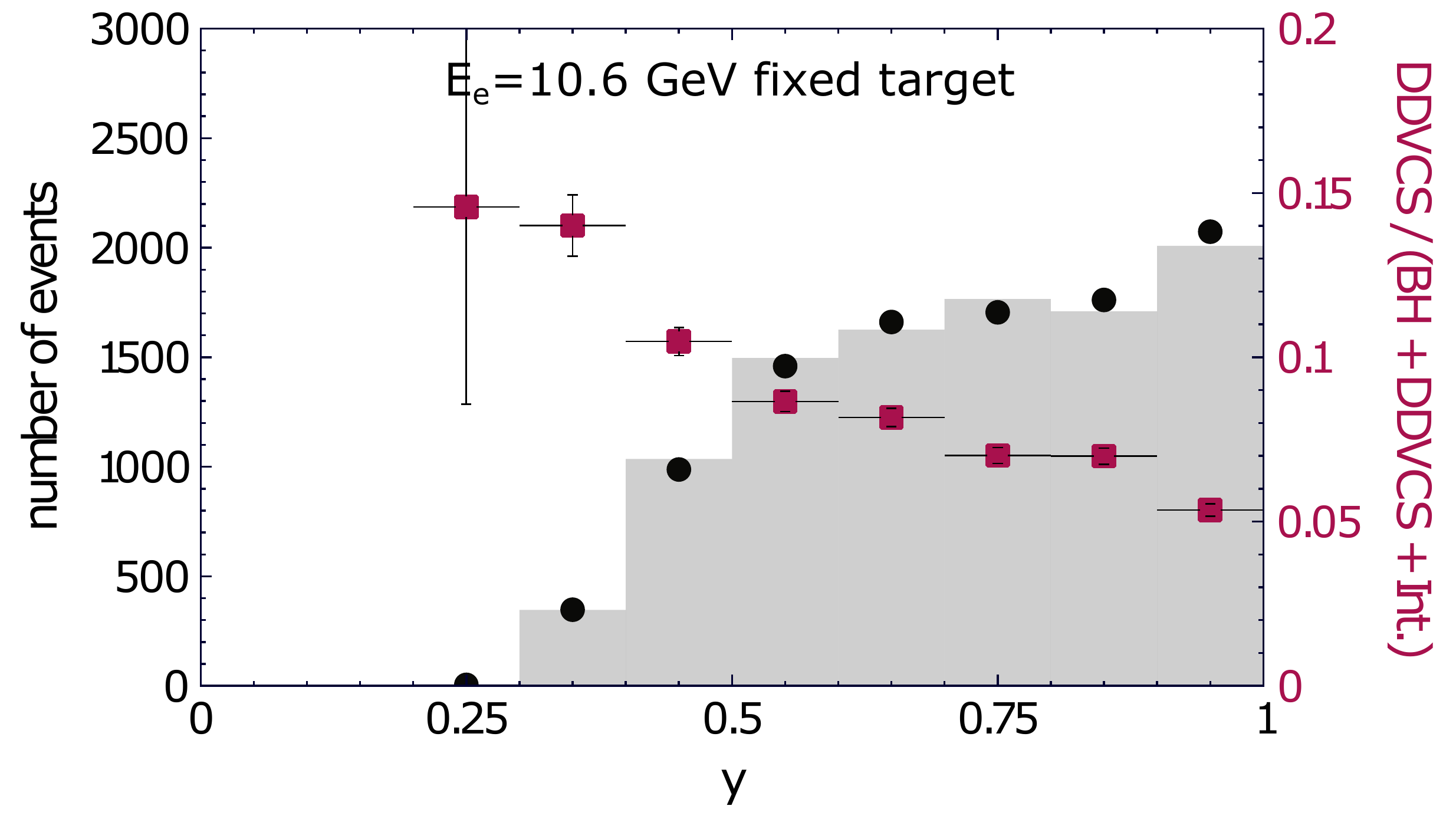}
    \includegraphics[width=0.45\textwidth]{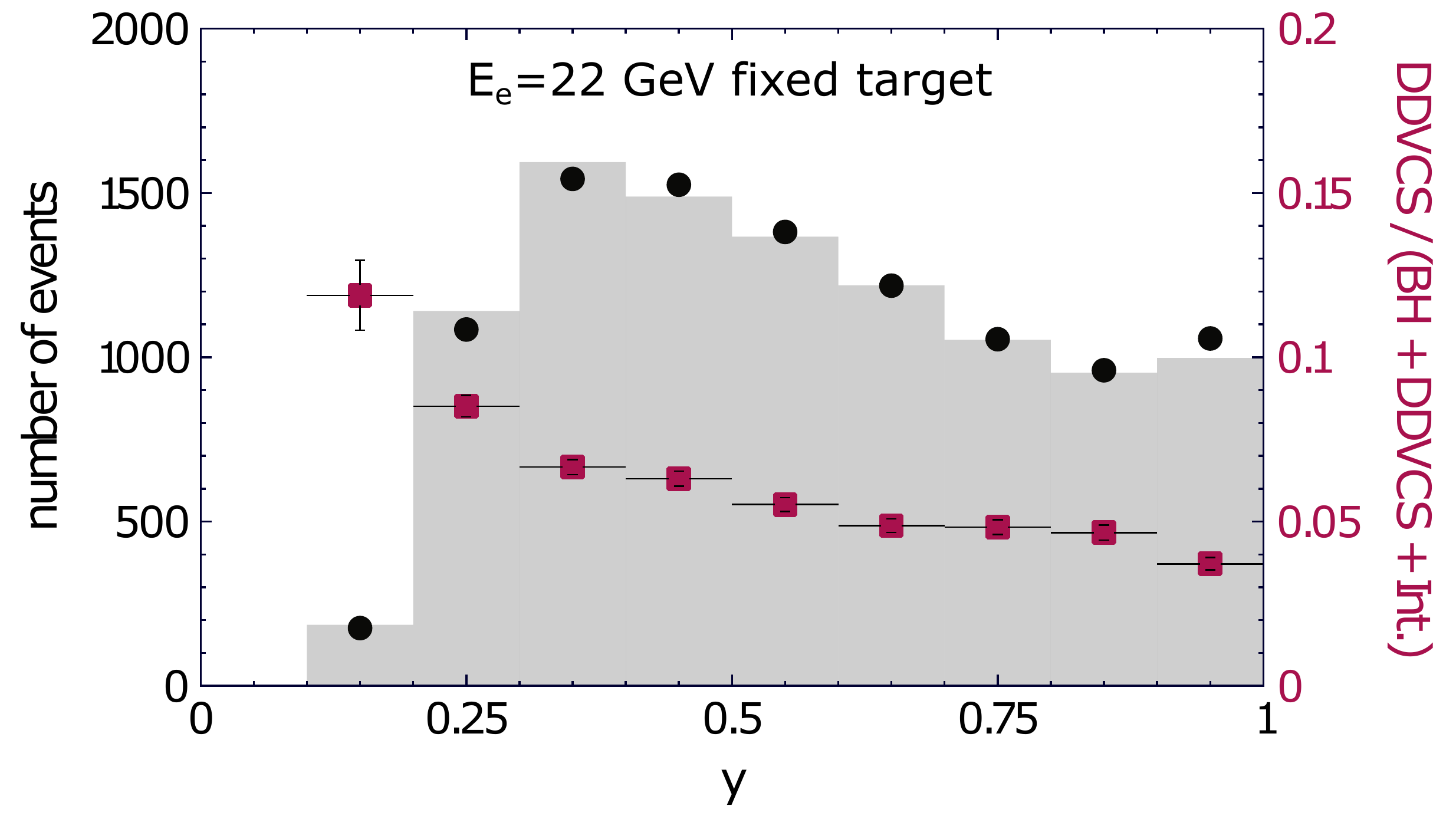}
    \includegraphics[width=0.45\textwidth]{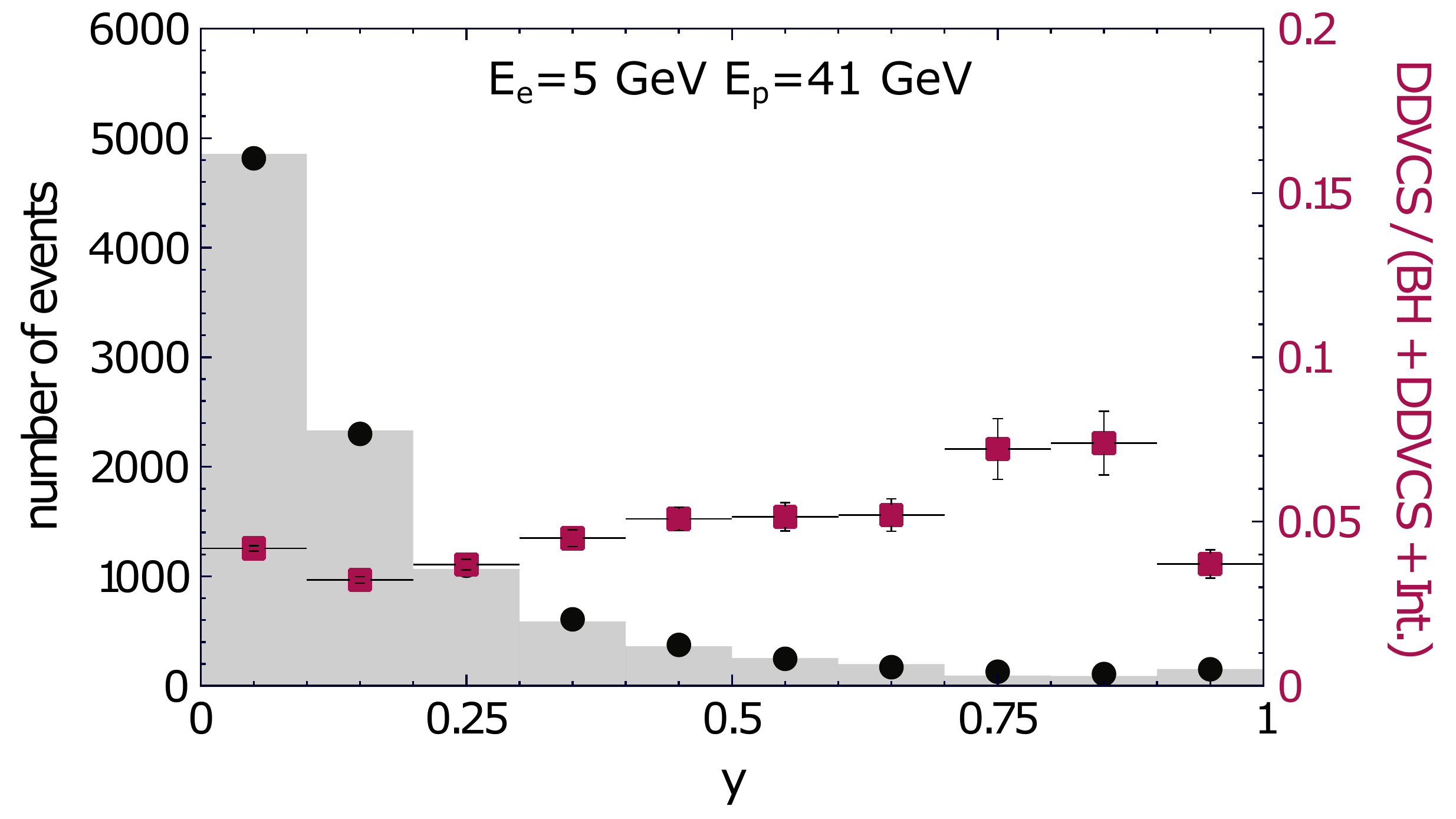}
    \includegraphics[width=0.45\textwidth]{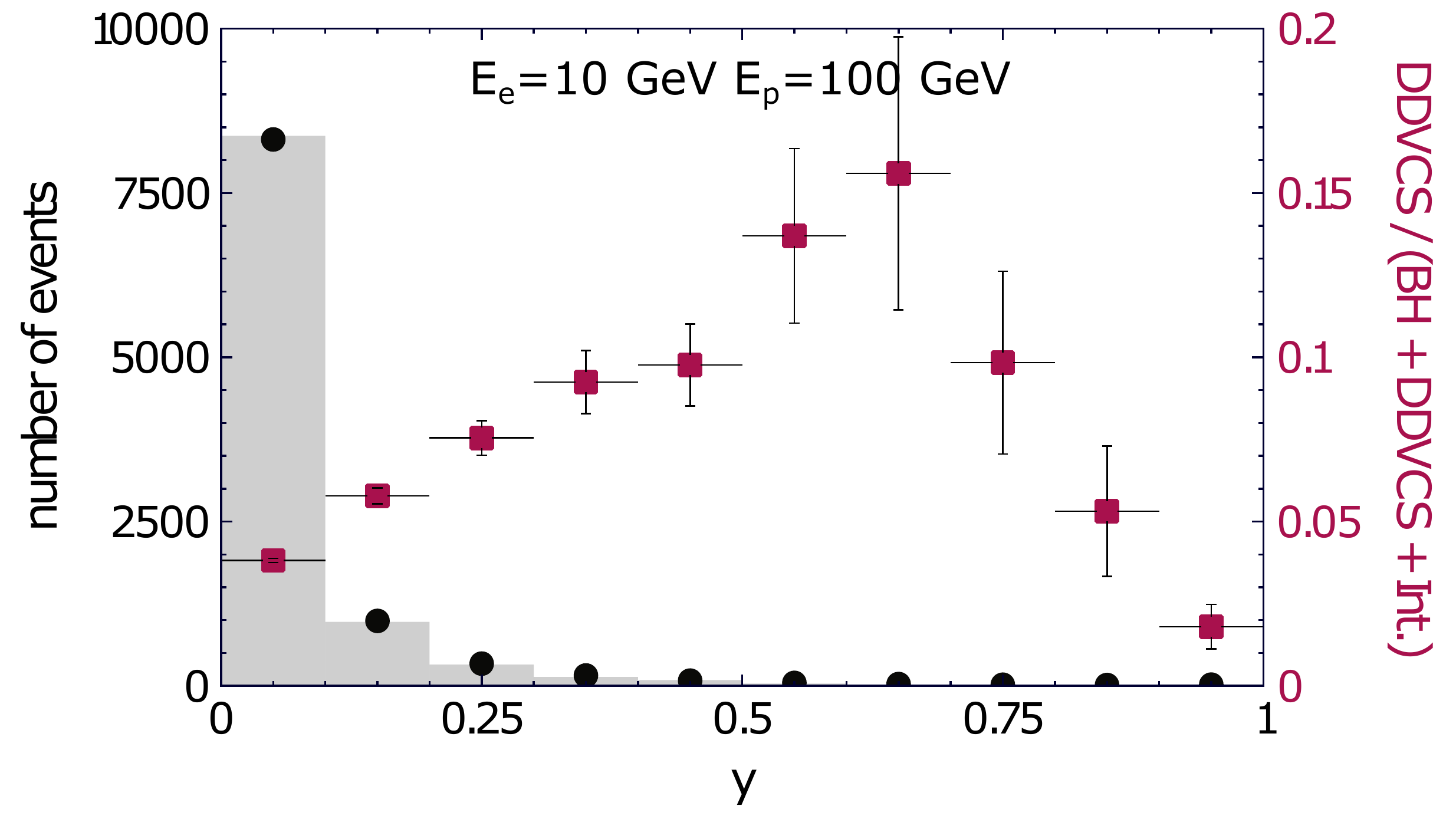}
    \caption{Distributions of Monte Carlo events as a function of the inelasticity variable $y$. Each distribution is populated by 10000 events generated for the beam energies specified in the plots. Extra kinematic conditions are specified in the text. Black circle markers and gray histograms correspond  to the left axes.   Reference values for Monte Carlo distributions obtained with  a direct integration of differential cross-section. The fraction of  events coming from the VCS sub-process with respect to all Monte Carlo events is indicated by red square markers corresponding to the right axes.}
    \label{figure:MCHist}
\end{figure}

The distribution of MC events we obtained as a function of $y$ is shown in Fig.~\ref{figure:MCHist}. The generation was done separately for four configurations of electron, $E_{e}$, and proton, $E_{p}$,  beam energies : i) $E_{e} = 10.6\ \mathrm{GeV}$ and fixed target, ii) $E_{e} = 22\ \mathrm{GeV}$ and fixed target, iii) $E_{e} = 5\ \mathrm{GeV}$ and $E_p = 41\ \mathrm{GeV}$, iv) $E_{e} = 10\ \mathrm{GeV}$ and $E_p = 100\ \mathrm{GeV}$; corresponding to JLab12, JLab20+ and EIC experiments. Additional conditions were used in the generation. The range of $Q^2$ variable was limited to $(0.15, 5)\ \mathrm{GeV}^2$. The lower value corresponds to the anticipated threshold for detection of scattered electrons, while the upper one is a reasonable limit for the observation of cross-section that is suppressed when $Q^2$ becomes large. The range of $Q'^2$ is $(2.25, 9)\ \mathrm{GeV}^2$ and it corresponds to the TCS analysis presented in Ref. \cite{CLAS:2021lky}. For $|t|$ we assumed $(0.1, 0.8)\ \mathrm{GeV}^2$ for JLab experiments and $(0.05, 1)\ \mathrm{GeV}^2$ for EIC ones. As for the angular dependencies, the ranges for $\phi$ and $\phiL$ angles are  $(0.1, 2\pi - 0.1)$, while for $\thetaL$ we have $(\pi/4, 3\pi/4)$. The limitations on angles help to suppress contributions coming from the BH sub-process. 

The total cross-section for the scattering (\ref{reaction}), including all  sub-processes, i.e.  BH, pure DDVCS and their interference, integrated in the aforementioned kinematic domain is tabulated in Table \ref{tab:MCCS}. In this table we also specify the integrated luminosity needed to record 10000 events presented in Fig.~\ref{figure:MCHist}, and the fraction of events recovered after cutting on the $y$ variable: $(0.1, 1)$ for JLab experiments and $(0.05, 1)$ for EIC ones.  

\begingroup
\begin{table}[!ht]
\begin{ruledtabular}
\begin{tabular}{@{}lcccccc@{}}
Experiment & Beam energies & Range of $|t|$ & $\sigma \rvert_{0<y<1}$ & $\mathcal{L}^{10\mathrm{k}}\rvert_{0<y<1}$ & $y_{\mathrm{min}}$ & $\sigma \rvert_{y_{\mathrm{min}} < y < 1} / \sigma \rvert_{0<y<1}$ \\ 
& $[\mathrm{GeV}]$ & $[\mathrm{GeV}^2]$ & $[\mathrm{pb}]$ & $[\mathrm{fb}^{-1}]$ & & \\[5pt] 
JLab12 & $E_{e} = 10.6$, $E_p = M$ & $(0.1, 0.8)$ & $0.14$ & $70$ & $0.1$ & $1$\\
JLab20+ & $E_{e} = 22$, $E_p = M$ & $(0.1, 0.8)$ & $0.46$ & $22$ & $0.1$ & $1$\\ 
EIC & $E_{e} = 5$, $E_p = 41$ & $(0.05, 1)$ & $3.9$ & $2.6$ & $0.05$ & $0.73$\\
EIC & $E_{e} = 10$, $E_p = 100$ & $(0.05, 1)$ & $4.7$ & $2.1$ & $0.05$ & $0.32$ \\
\end{tabular}
\end{ruledtabular}
\caption{Total DDVCS cross-section including all sub-processes, $\sigma \rvert_{0<y<1}$, obtained for given beam energies under the following conditions: $y \in (0, 1)$, $Q^2 \in (0.15, 5)\ \mathrm{GeV}^2$, $Q'^2 \in (2.25, 9)\ \mathrm{GeV}^2$, $\phi, \phiL \in(0.1, 2\pi - 0.1)$, $\thetaL \in (\pi/4, 3\pi/4)$ and $|t|$ range specified in 3th column. Corresponding integrated luminosity required to obtain 10000 events is denoted by $\mathcal{L}^{10\mathrm{k}}\rvert_{0<y<1}$. Fraction of events left after restricting the range of $y$ to $(y_{\mathrm{min}}, 1)$ is given in the last column.}
\label{tab:MCCS}
\end{table}
\endgroup

In Fig.~\ref{figure:MCHist} we also show the expected number of events, coming from a direct seven-fold integration of cross-section (\ref{xsec}) in the aforementioned kinematic domain and limits of $y$ specified by a given bin of the histogram. No free normalization factor is used here: integrated cross-section is multiplied by the luminosity given by EpIC. The comparison between obtained values and MC samples proves the correctness of the generation. 

An additional quantity shown in Fig.~\ref{figure:MCHist} is the fraction of pure DDVCS sub-process in the sample. As expected, this fraction is small, which stresses the need for measuring observables sensitive to the interference in the latter case. 

\begin{figure}
    \centering
    \includegraphics[width=0.45\textwidth]{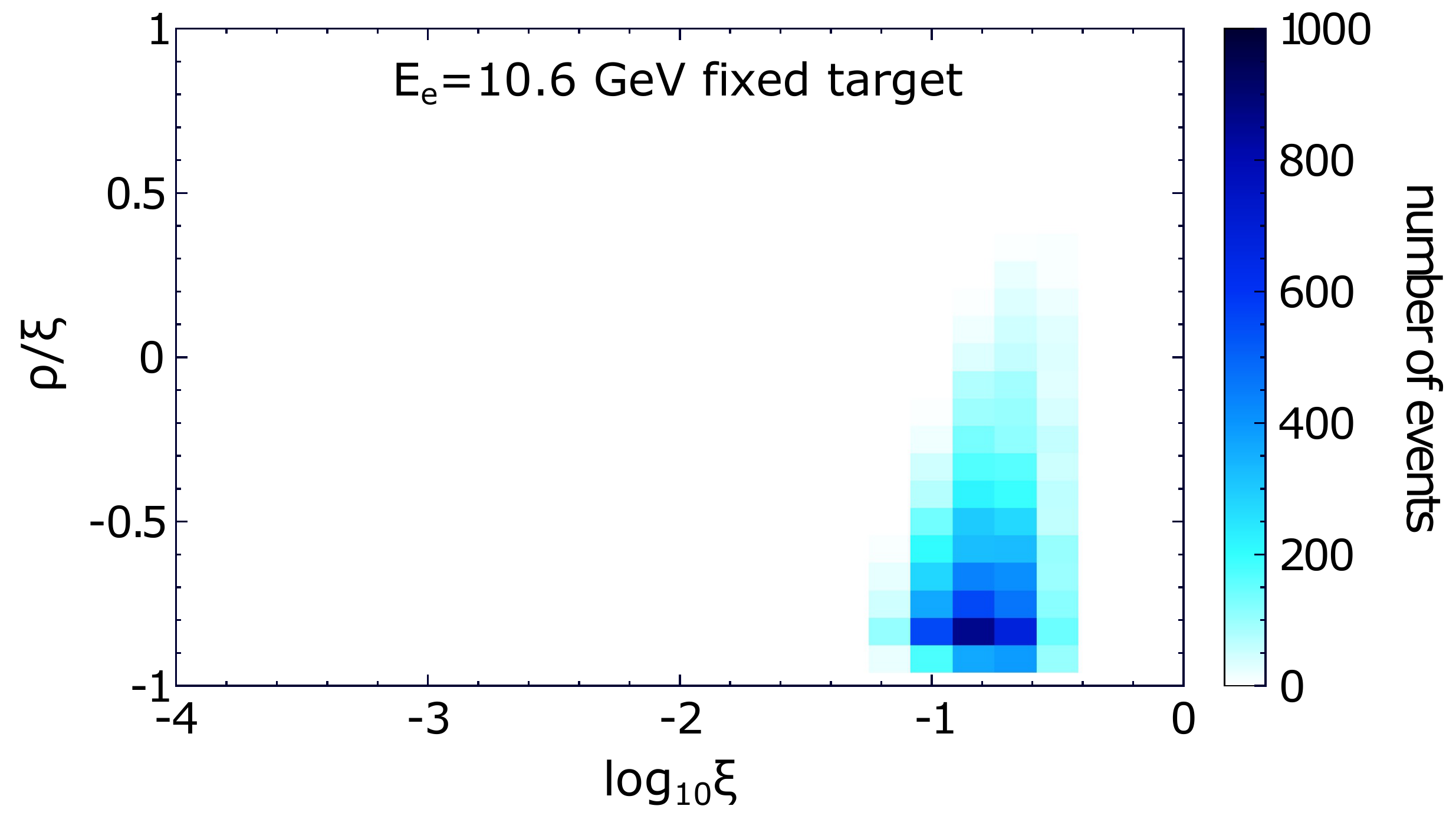}
    \includegraphics[width=0.45\textwidth]{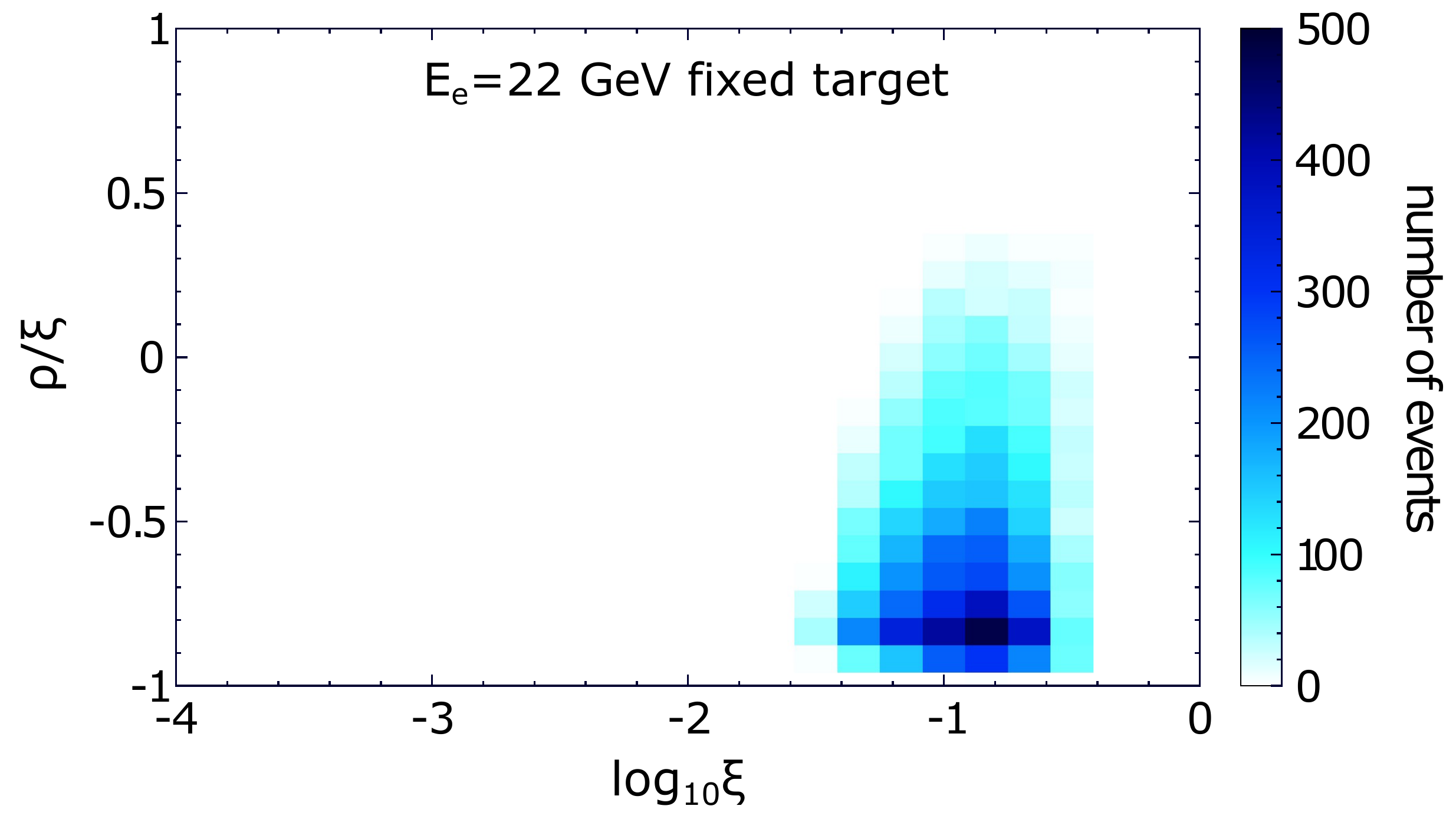}
    \includegraphics[width=0.45\textwidth]{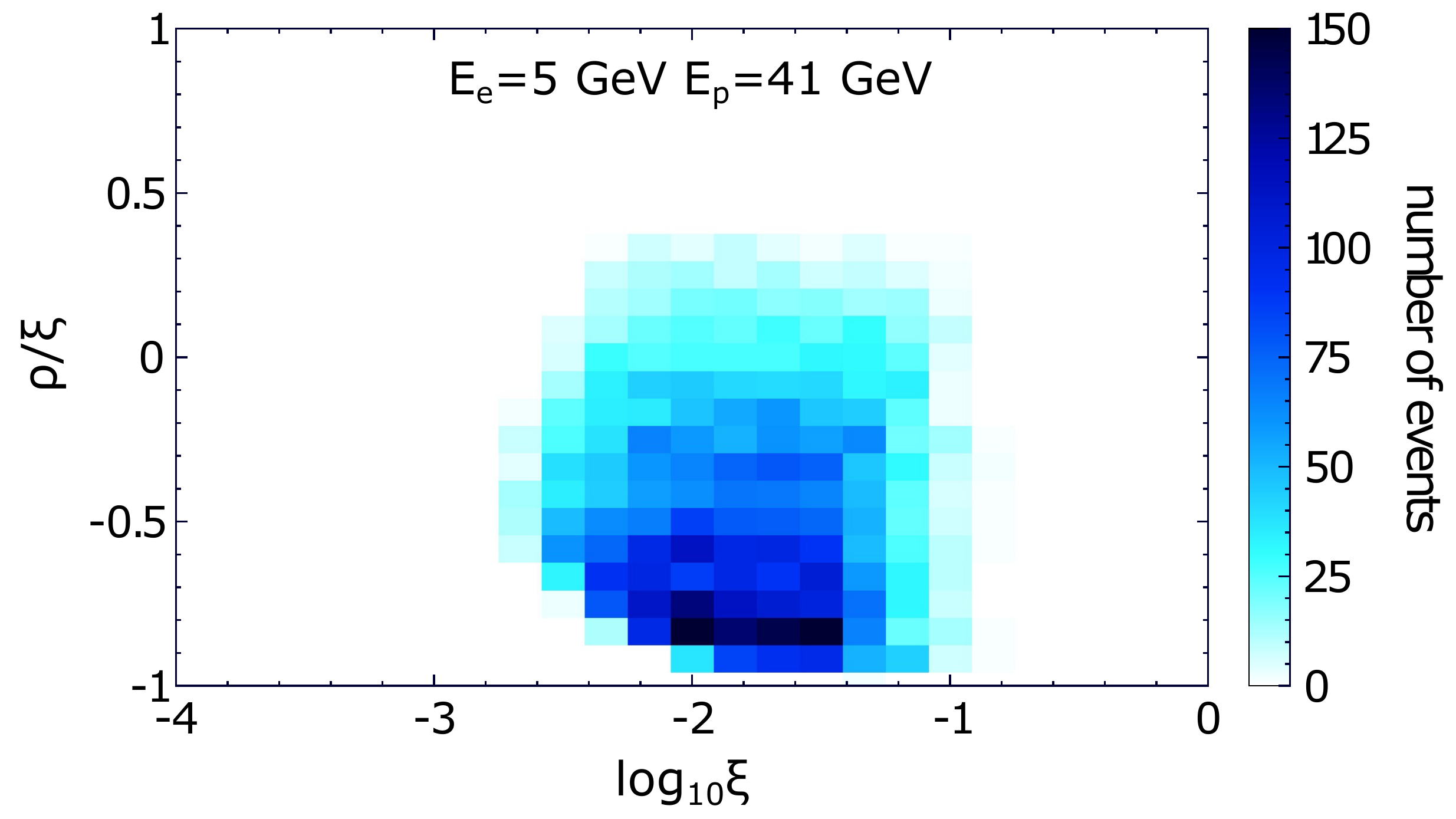}
    \includegraphics[width=0.45\textwidth]{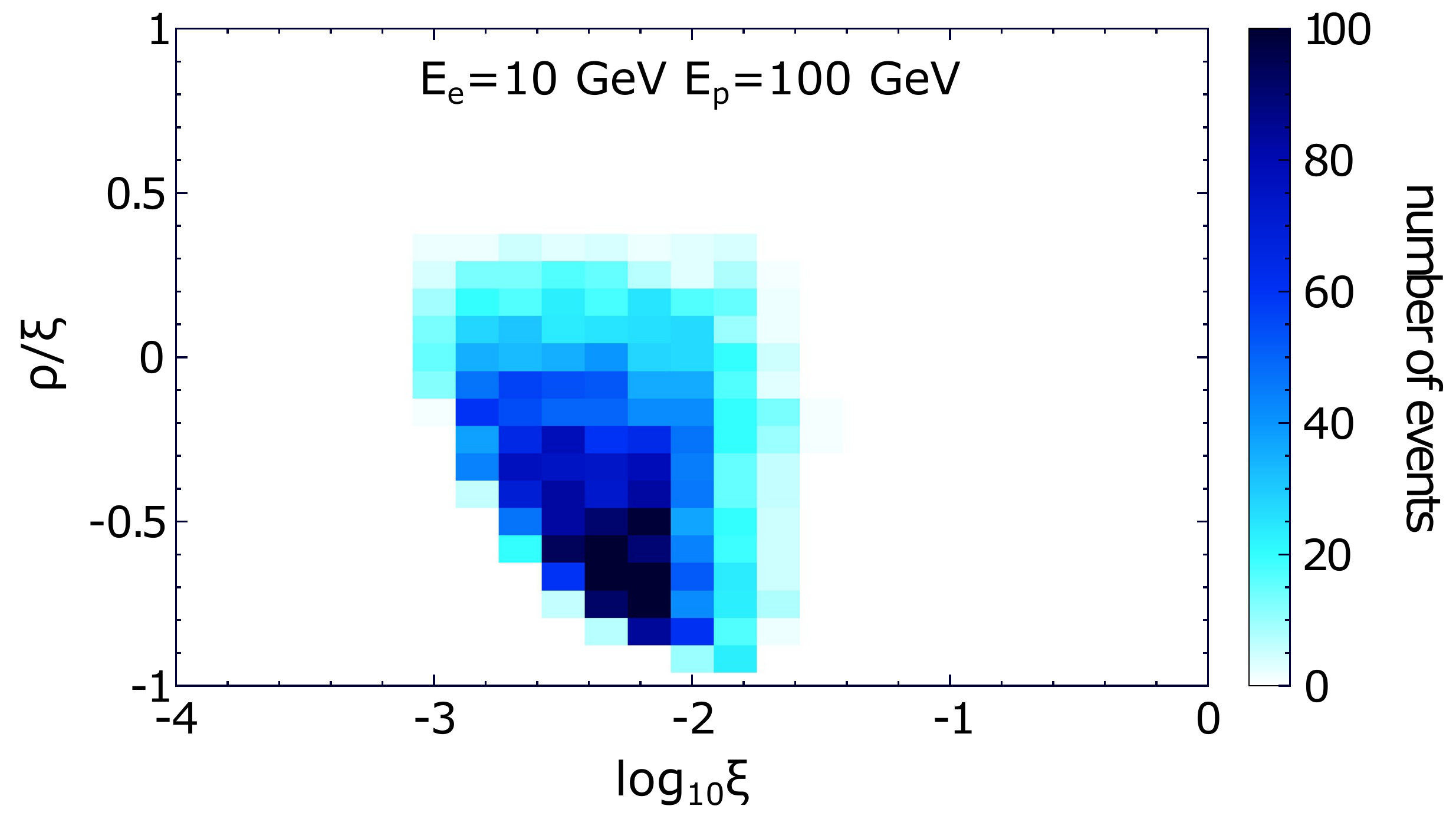}
    \caption{Distribution of Monte Carlo events as a function of the skewness variable $\xi$ and the relative value of generalized Bj\"orken variable $\rho$. Each distribution is populated by 10000 events generated for the DDVCS sub-process at beam energies specified in the plot. Extra kinematical conditions, including cuts on the $y$ variable, are specified in the text.}
    \label{figure:MCxiVsRho}
\end{figure}

Additional information is provided in Fig. \ref{figure:MCxiVsRho}, where we show how pure DDVCS events populate the $\xi$ vs.~$\rho$ phase-space. Clearly, the $\xi \neq |\rho|$ domain is probed, proving the importance of the DDVCS reaction in the reconstruction of GPDs from experimental data. We note that because of the ranges choosen for $Q^2$ and $Q'^2$ we typically probe the negative, ``more'' time-like, domain~of~$\rho$. 

\section{Conclusion}
\label{sectConc}

In this article we revisit the phenomenology of DDVCS process, which in the future may become one of the main sources of information on GPDs in the unexplored $\xi \neq x$ domain. The computation of BH and DDVCS amplitudes based on Kleiss-Stirling techniques allows us to express the results in the form of easy to implement complex quantities. The article also includes predictions for a new generation of experiments, including MC studies. 

Our preliminary impact studies provide promising results in terms of DDVCS measurability and encourage further analyses for future experimental facilities. The values of cross-sections integrated in reasonable kinematic ranges accessible to experiments (including $y$-cuts) are: $0.14~\mathrm{pb}$ for JLab12, $0.46~\mathrm{pb}$ for JLab20+, $2.8~\mathrm{pb}$ and $1.5~\mathrm{pb}$ for EIC experiments running with $5~\mathrm{GeV}\times41~\mathrm{GeV}$ and $10~\mathrm{GeV}\times100~\mathrm{GeV}$ beam energy configurations, respectively. We note that our study does not include any detector effects. We stress that a successful DDVCS programme will not only depend on high luminosity accessible to future machines, but also on effective muon reconstruction and possibility of running experiments with various beam and target polarization states. 
Our analysis should be complemented with NLO expressions of DDVCS CFFs \cite{Pire:2011st}. These NLO contributions include amplitudes proportional to gluon GPDs. Since the inclusion of these corrections turn out to be phenomenologically important in both DVCS and TCS cases \cite{Moutarde:2013qs}, we can anticipate that they will be also non-negligible in the DDVCS case. Moreover, they give access to the elusive gluon transversity GPDs \cite{Belitsky:2000jk}. The inclusion of kinematical higher-twist corrections, which has been demonstrated for the DVCS case in Ref.~\cite{Braun:2014sta}, should also be performed for DDVCS. We leave these improvements for a future study.
\\
\paragraph*{Acknowledgements.}
\noindent
The authors would like to thank Sebastian Alvarado, Lech Szymanowski and Eric Voutier for fruitful discussions. The works of J.W. are  supported by the grant 2017/26/M/ST2/01074 of the National Science Center (NCN), Poland. Development of EpIC Monte Carlo generator by P.S. was supported by the grant 2019/35/D/ST2/00272 of the National Science Centre (NCN), Poland. This work is also partly supported by the COPIN-IN2P3 and by the European Union’s Horizon 2020 research and innovation programme under grant agreement No 824093. The works of V.M.F.~are supported by PRELUDIUM grant 2021/41/N/ST2/00310 of NCN.

\bibliography{DDVCS_paper.bib}

\end{document}